\documentclass[11pt]{article}
\usepackage{harvard,graphicx}
\usepackage{epsfig}
\usepackage{float}
\usepackage{lscape}
\usepackage{amsfonts}
\usepackage{amsmath}
\usepackage[latin9]{inputenc}
\usepackage{amsmath}
\usepackage{amssymb}
\usepackage{graphicx}
\usepackage[unicode=true,
 bookmarks=false,
 breaklinks=false,pdfborder={0 0 1},backref=none,colorlinks=false]{hyperref}
\usepackage{epstopdf,subfig}
\usepackage{amsfonts}
\usepackage{harvard}
\usepackage{epsfig}
\usepackage{float}
\usepackage{threeparttable,booktabs}
\usepackage{placeins}
\usepackage[onehalfspacing]{setspace}

\makeatletter

\renewcommand{\cite}{\citeasnoun}
\input{tcilatex}
\begin{document}

\begin{spacing}{1.25}
\title{The Informativeness of Estimation Moments\thanks{%
We gratefully acknowledge financial support from
the National Science Foundation (grant numbers SES-1530741 and SES-1824131),
the Danish Council for Independent Research in Social Sciences (FSE, grant
no. 4091-00040), the ESRC through the Centre for Microdata Methods and
Practice (RES-589-28-0001) and the ERC (SG338187). The activities of Center
for Economic Behavior and Inequality (CEBI) are financed by a grant from the
Danish National Research Foundation. We thank Sharada Dharmasankar, the coeditor and three anonymous referees for constructive comments and suggestions.
A previous version of this paper was distributed under the title ``Sensitivity of Estimation Precision to
Moments with an Application to a Model of Joint Retirement Planning
of Couples.''
}}
\author{Bo E. Honor\'{e}\thanks{%
 Department of Economics, Princeton University and The Dale T. Mortensen
Centre at the University of Aarhus. E-mail: \href{mailto:honore@princeton.edu}%
{honore@princeton.edu}} \and Thomas H. J{\o}rgensen\thanks{%
 Center for Economic Behavior and Inequality (CEBI), Department of Economics,
University of Copenhagen. E-mail: \href{mailto:thomas.h.jorgensen@econ.ku.dk}%
{thomas.h.jorgensen@econ.ku.dk}. Webpage: \href{www.tjeconomics.com}{%
www.tjeconomics.com}.} \and \'Aureo de Paula\thanks{%
Department of Economics, University College London, C\emph{e}MMAP and IFS, London, UK. E-mail: \href{mailto:apaula@ucl.ac.uk}%
{apaula@ucl.ac.uk}}}

\maketitle

\begin{abstract}

\noindent This paper introduces measures for how each moment contributes to the precision of parameter estimates in GMM settings.
For example, one of the measures asks what would happen to the variance of the parameter estimates if a particular moment was dropped from the estimation.
The measures are all easy to compute.  We illustrate the usefulness of the measures through two simple examples as well as an application to a model of joint retirement planning of couples. We estimate the model using the UK-BHPS, and we find evidence of complementarities in leisure. Our sensitivity measures illustrate that the estimate of the complementarity is primarily informed by the distribution of differences in planned retirement dates. The estimated econometric model can be interpreted as a bivariate ordered choice model that allows for simultaneity.  This makes the model potentially useful in other applications.


\noindent JEL\ C\textsc{ode}: C01, C13, C35, J32.

\end{abstract}
\end{spacing}

\thispagestyle{empty} \newpage{} \pagenumbering{arabic} \setcounter{page}{1}

\section{Introduction \label{Sec: Introduction}}

Indirect inference and other nonlinear GMM estimators are used extensively
in empirical research. These estimators are, however, sometimes seen as
black boxes. It can be difficult to understand exactly what features of the
data are informative about which parameters, and how sensitive parameter
estimates are to  moments  included in the objective function.

In this paper, we provide simple and easy-to-compute measures that can
indicate how altering the moments used in estimation affects the precision
of parameter estimates. Informally, we think of these as measures of how
informative each moment is about a particular parameter. More precisely, we
provide measures of the effect on asymptotic standard errors from \emph{i)}
a marginal increase in the noise associated with a moment, \emph{ii)}
completely removing a (set of) moments from estimation, and \emph{iii)} a
marginal increase in the weight put on a moment.

The measures are derived from the asymptotic distribution of the class of
GMM-type estimators considered here and are, for the most part, based on
derivatives of the asymptotic covariance matrix. The measures are almost
costless to calculate because most of the required quantities are already
constructed when calculating asymptotic standard errors. Furthermore, the
measures have straightforward interpretations if scaled in a meaningful way.

There is a growing literature investigating sensitivity of estimators in
economics. Recently, for example, \cite%
{AndrewsGentzkowShapiro2017_sensitivity} proposed a measure to inform
researchers on the sensitivity of the asymptotic bias in estimators to
misspecification of moments included in the estimation function. We note
that their measure is also related to the change in the asymptotic variance
from a marginal change in the included moments, which inspired our proposed
alternative measures. While we focus on the precision of the parameter
estimates, more recently \cite{ArmstrongKolesar2018} and \cite%
{BonhommeWeidner2018} have also studied local misspecification. \cite%
{ChristensenConnault2019} studied global misspecification.

We illustrate the applicability of our measures through two simple examples
and an empirical application. The two examples are a binary outcome probit model and a
proportional hazards Weibull duration model with time-varying covariates.
The application is a simple structural model of joint retirement planning of
dual-earner households. The model is founded in utility maximization with
household bargaining, but can also be interpreted as a bivariate ordered
choice model that allows for simultaneity. The parameters of the model are
most easily estimated by indirect inference, but the complexity of the model
makes it difficult to understand the link between the data and the parameter
estimates.

While a growing empirical literature has established that dual earner
households tend to retire simultaneously or in quick succession in age,%
\footnote{%
See e.g. \cite{Hurd1990}; \cite{Blau1998}; \cite{GustmanSteinmeier2000};
\cite{GustmanSteinmeier2004}; \cite{Coile2004}; \cite{AnChristensenGupta2004}%
; \cite{Jia2005}; \cite{BlauGilleskie2006}; \cite{vanderKlaauwWolpin2008};
\cite{BanksBlundellCasanova2010}; \cite{Casanova2010} and \cite%
{HonorePaula2018}.} the empirical evidence of joint retirement \textit{%
planning} of couples is much more scarce and with ambiguous findings.%
\footnote{%
See \cite{PientaHayward2002}; \cite{MoenHuangPlassmannDentinger2006}; and
\cite{deGripFouargeMontizaan2013}.} We contribute to this literature by
estimating a structural model of dual-earner retirement planning using
indirect inference and prospective retirement planning questions in the
British Household Panel Survey (BHPS). Our estimation results support the
notion of leisure complementarities in retirement. Our proposed sensitivity
measures confirm the intuition that the parameter estimate measuring leisure
complementarities in the model is sensitive to the distribution of the
difference in the year of planned retirement between household members.

The remaining paper is organized as follows. In Section \ref{Sec:
Sensitivity Measures}, we present the sensitivity measures and show examples
of their use in Section \ref{Sec: Examples}. In Section \ref{Sec:
Application}, we apply our measures to a novel model of dual earner
retirement planning before concluding with final remarks in Section \ref%
{Sec: Conclusion}.

\section{Framework and Sensitivity Measures\label{Sec: Sensitivity Measures}}

Indirect inference and other nonlinear GMM estimators are sometimes seen as
black boxes where it can be difficult to understand exactly what features of
the data are informative about which parameters. In this section, we review
and introduce a number of measures that are meant to provide information
about this.

To fix ideas, consider a set of moment conditions $E\left[ f\left(
x_{i},\theta _{0}\right) \right] =0$, where $x_{i}$ is data for observation $%
i$ and it is assumed that this defines a unique $\theta _{0}$. The
generalized method of moments (GMM) estimator of $\theta _{0}$ is $\widehat{%
\theta }=\arg \min_{\theta }\left( \frac{1}{n}\sum_{i=1}^{n}f\left(
x_{i},\theta \right) \right) ^{\prime }W_{n}\left( \frac{1}{n}%
\sum_{i=1}^{n}f\left( x_{i},\theta \right) \right) $, where $W_{n}$ is a
symmetric, positive definite matrix. While some of the measures below also
apply to just-identified models, we focus here on over-identified models
where the number of moments are larger than the number of parameters in $%
\theta$, and the weighting matrix thus plays a role.

Subject to the standard regularity conditions, the derivation of the
asymptotic distribution of $\widehat{\theta }$ gives
\begin{equation}
\widehat{\theta }=\theta _{0}-\left( G^{\prime }WG\right) ^{-1}G^{\prime
}W\left( \frac{1}{n}\sum_{i=1}^{n}f\left( x_{i},\theta _{0}\right) \right)
+o_{p}\left( n^{-1/2}\right) ,  \label{expansion}
\end{equation}%
where $G=E\left[ \frac{\partial f\left( x_{i},\theta _{0}\right) }{\partial
\theta }\right] $ and $W$ is the limit of $W_{n}$. See \cite{Hansen1992}.
The limiting distribution of the GMM estimator is
\begin{equation*}
\sqrt{n}\left( \widehat{\theta }-\theta _{0}\right) \overset{d}{%
\longrightarrow }N\left( 0,\Sigma \right) ,
\end{equation*}%
where
\begin{equation*}
\Sigma =\left( G^{\prime }WG\right) ^{-1}G^{\prime }WSWG\left( G^{\prime
}WG\right) ^{-1}
\end{equation*}%
and $S=V\left[ f\left( x_{i},\theta _{0}\right) \right] $ under random
sampling. If we use the optimal weighting matrix, $W=S^{-1}$, the asymptotic
covariance collapses to
\begin{equation*}
\Sigma _{opt}=(G^{\prime }S^{-1}G)^{-1}.
\end{equation*}%
Intuitively, when there is little sampling variability in the moment
functions, $f$, $S$ will be small. $G$ is larger if the moment condition is
more sensitive to perturbations in the parameter. Both of these contribute
to the precision of the estimates as the proposed measures highlight.

\cite{AndrewsGentzkowShapiro2017_sensitivity} proposed the sensitivity
measure
\begin{equation*}
M_{1}=-(G^{\prime }WG)^{-1}G^{\prime }W.
\end{equation*}%
It is clear from (\ref{expansion}) that $M_{1}$ provides the mapping from
moment misspecification of the type $E\left[ f\left( x_{i},\theta
_{0}\right) \right] =\rho \neq 0$ into parameter biases for small $\rho $.
Alternatively, by noting that $\Sigma =M_{1}SM_{1}^{\prime }$, $M_{1}$ tells
us how additional noise in each of the sample moments $\frac{1}{n}%
\sum_{i=1}^{n}f\left( x_{i},\theta _{0}\right) $ would result in additional
noise in each element of $\widehat{\theta }$. This is what motivates our
alternative measures that address the sensitivity of estimation \textit{%
precision} to each moment.

The proposed measures are intended to complement the measure of sensitivity
to misspecification proposed by \cite{AndrewsGentzkowShapiro2017_sensitivity}%
. Like $M_{1}$, our measures are matrices where the $\left( j,k\right) $'th
element provides an answer to how the precision of the $j$'th element of $%
\widehat{\theta }\ $\ depends on the $k$'th moment.

Our first measure asks the hypothetical question: How much precision would
we lose if the $k$'th moment is subject to a little additional noise? This
measure is formally defined as
\begin{equation*}
M_{2,k}\equiv \frac{\partial \Sigma _{opt}}{\partial S^{(kk)}}=\Sigma
_{opt}(G^{\prime }S^{-1}O_{kk}S^{-1}G)\Sigma _{opt},
\end{equation*}%
where $O_{kk}$ is a matrix with 1 in the $(k,k)$ element and zero elsewhere.
This measure assumes that the optimal weighting matrix is used and updated.
Alternatively, we could ask the same question keeping the (possibly
non-optimal) weighting matrix unchanged. This measure is
\begin{equation*}
M_{3,k}\equiv \frac{\partial \Sigma }{\partial S^{(kk)}}=(G^{\prime
}WG)^{-1}G^{\prime }WO_{kk}WG(G^{\prime }WG)^{-1}=M_{1}O_{kk}M_{1}^{\prime }.
\end{equation*}%
The difference between $M_{2,k}$ and $M_{3,k}$ is that the former evaluates
the potential information in each moment while the latter evaluates the
information actually used in the estimation. With efficient GMM (so $W=S^{-1}
$), $M_{3}$ equals $M_{2}$. This is also true in the just-identified case
where the number of moments equals the number of parameters to be estimated.

Related to $M_{2,k}$, we could consider the change in the asymptotic
variance from completely excluding the $k$'th moment,
\begin{equation*}
M_{4,k}\equiv \tilde{\Sigma}_{k}-\Sigma ,
\end{equation*}%
where
\begin{align*}
\tilde{\Sigma}_{k}& =(G^{\prime }\tilde{W}_{k}G)^{-1}G^{\prime }\tilde{W}%
_{k}S\tilde{W}_{k}G(G^{\prime }\tilde{W}_{k}G)^{-1} \\
\tilde{W}_{k}& =W\odot (\iota _{k}\iota _{k}^{\prime }).
\end{align*}%
Here $\odot $ denotes element-wise multiplication and $\iota _{k}$ is a $%
J\times 1$ vector with ones in all elements except the $k$'th element, which
is zero. $M_{4,k}$ leaves the weighting matrix on the remaining moments
unchanged after we have excluded the $k$'th moment.

We note that this measure assumes that the parameter vector is identified
after the $k$'th moment has been excluded. Specifically, $(G^{\prime }\tilde{%
W}_{k}G)$ needs to have full rank. Importantly, this means that the original
model has to be over-identified in the sense that it has more moments than
parameters. In practice, $G$ has to be estimated, and violations of the full
rank assumption will result in $(\widehat{G}^{\prime }\tilde{W}_{k}\widehat{G%
})$ being close to singular. Extremely large values in the estimate of $%
M_{4,k}$ therefore suggest that the model is not point-identified when the $k
$'th moment is excluded. This can happen even if the original model was
over-identified.

Alternatively, one could also consider measures that adjust the weighting
matrix. For example, one could consider a measure that compares the
precision of the optimal GMM\ estimator that uses all moments to the optimal
GMM\ estimator that excludes that $k$'th moment,%
\begin{equation*}
M_{5,k}=(G_{-k}^{\prime }S_{-k}^{-1}G_{-k})^{-1}-(G^{\prime }S^{-1}G)^{-1},
\end{equation*}%
where $G_{-k}$ is the same as matrix $G$ except that the $k$'th row has been
removed, and $S_{-k}$ is $S$ with the $k$'th row and column removed. This
measure also assumes that the parameter vector is identified after the $k$%
'th moment has been excluded, and it implicitly assumes that the original number of 
moment conditions exceeds the number of parameters to be estimated.

 $M_{4}\ $and $M_{5}$ can also be used to gauge the sensitivity
of the estimator to a set of moments. This is potentially useful in cases
where one can group moments in some natural way. One can then address
the question of how much of the precision in an estimator would be lost if
one did not use one of the groups of moments. For example, \cite%
{GayleShephard2019} talks about five sets of moments (in their online
appendix), and \cite{HonorePaula2018} get their moments from the estimation
of four different auxiliary reduced form models. A reparameterization of
those reduced form models would lead to moment conditions which are
(asymptotically) linear combinations of the original moment conditions. In
that case, it might be useful to construct a measure that reflects giving
zero weight to all the moments that come from a specific auxiliary model.
This approach would be application-specific, and we therefore do not pursue
it in this paper.

Our final measure addresses the question: How would the precision of our
estimates change if we slightly increased the weight put on the $k$th
moment? This measure is formally defined as the derivative
\begin{align*}
M_{6,k}\equiv \frac{\partial \Sigma }{\partial W^{\left( k,k\right) }}=-&
(G^{\prime }WG)^{-1}(G^{\prime }O_{kk}G)\Sigma +(G^{\prime
}WG)^{-1}G^{\prime }O_{kk}SWG(G^{\prime }WG)^{-1} \\
+& (G^{\prime }WG)^{-1}G^{\prime }WSO_{kk}G(G^{\prime }WG)^{-1}-\Sigma
(G^{\prime }O_{kk}G)(G^{\prime }WG)^{-1}.
\end{align*}%
We do not think of $M_{6,k}$ as a measure of moment sensitivity, but rather
as a measure of how close the chosen weighting matrix is to being optimal. $%
M_{6,k}$ will be 0 when $W$ is the optimal weighting matrix. It will
also be 0 in the just-identified case, where the number of moments equals the
number of parameters to be estimated.

These measures are not invariant to scale of the included moments in $%
f(\cdot )$. One approach, which we take, is to report scaled measures.
Concretely, we report the sensitivity of the $j$'th parameter to the $k$'th
moment as
\begin{align*}
\mathcal{E}_{2}^{(j,k)}& =M_{2}^{(j,k)}\frac{S^{(k,k)}}{\Sigma _{opt}^{(j,j)}%
} \\
\mathcal{E}_{3}^{(j,k)}& =M_{3}^{(j,k)}\frac{S^{(k,k)}}{\Sigma ^{(j,j)}} \\
\mathcal{E}_{4}^{(j,k)}& =M_{4}^{(j,k)}\frac{1}{\Sigma ^{(j,j)}} \\
\mathcal{E}_{5}^{(j,k)}& =M_{5}^{(j,k)}\frac{1}{\Sigma _{opt}^{(j,j)}} \\
\mathcal{E}_{6}^{(j,k)}& =M_{6}^{(j,k)}\frac{W^{(k,k)}}{\Sigma ^{(j,j)}}.
\end{align*}%
Note that $\mathcal{E}_{2}^{(j,k)}$, $\mathcal{E}_{3}^{(j,k)}$ and $\mathcal{%
E}_{6}^{(j,k)}$ are elasticities whereas $\mathcal{E}_{4}^{(j,k)}$ and $%
\mathcal{E}_{5}^{(j,k)}$ are the relative changes in the asymptotic variance
compared to the baseline with all moments included.

\section{Examples\label{Sec: Examples}}

In this section, we illustrate the use of our proposed measures through two
concrete examples. The first example is a simple binary choice probit model
and the second example is a proportional hazards duration model. The first
example is chosen because it is a case where one would have a strong
prior about which moments matter. The second example, on the other hand, is
complicated enough that this is not obvious.

For both examples, we use both the optimal weighting matrix and a diagonal
weighting matrix with the inverse of the moment variances on the diagonal.
We chose the latter non-optimal weighting matrix because it is very common
in empirical applications.\footnote{%
There are many examples of this. This includes \cite%
{EisenhauerHeckmanMosso2014} and \cite{GayleShephard2019} to name two. The
motivation stems from \cite{AltonjiSegal1996} who show that the optimal
weighting matrix can have quite poor finite sample properties. They suggest
equally weighted moments (i.e., $W=I$) as an alternative. Of course, using
equal weights will not be invariant to changes in units (or other
rescaling), which explains the practice we have adopted.}

\subsection{Example 1: Method of Moments Estimation of a Probit Model\label%
{Probit}}

We first consider a simple probit model
\begin{align*}
y_{i}& =\left\{
\begin{array}{ll}
0 & \text{if }y_{i}^{\ast }>0 \\
1 & \text{else}%
\end{array}%
\right. \\
y_{i}^{\ast }& =\beta _{0}+\beta _{1}x_{1,i}+\beta _{2}x_{2,i}+\varepsilon
_{i}
\end{align*}%
where $\left( x_{1,i},x_{2,i}\right) $ has a bivariate normal distribution
with means equal to 0, variance 1 and correlation 0.5. $\varepsilon _{i}$ is
independent of $\left( x_{1,i},x_{2,i}\right) $ and distributed according to
a standard normal. We set $\beta _{0}=\beta _{1}=\beta _{2}=1/\sqrt{3}$.
This makes $V\left[ \beta _{0}+\beta _{1}x_{1,i}+\beta _{2}x_{2,i}\right] =1$
and $P\left( y_{i}=1\right) =0.66$.

We consider the asymptotic distribution of a moment-based estimator of $%
\theta _{0}=(\beta _{0},\beta _{1},\beta _{2})$ solving
\begin{equation*}
\hat{\theta}=\arg \min_{\theta }g(\theta )^{\prime }Wg(\theta ),
\end{equation*}%
where we use the six moments
\begin{equation*}
\left(
\begin{array}{cccccc}
E[e\left( \theta \right) ] & E[e\left( \theta \right) x_{1}] & E[e\left(
\theta \right) x_{2}] & E[e\left( \theta \right) x_{1}^{2}] & E[e\left(
\theta \right) x_{1}x_{2}] & E[e\left( \theta \right) x_{2}^{2}]%
\end{array}%
\right) ^{\prime }
\end{equation*}%
and $e_{i}\left( \theta \right) =y_{i}-\Phi \left( \beta _{0}+\beta
_{1}x_{1,i}+\beta _{2}x_{2,i}\right) $. In the corresponding logit model,
the first three moments correspond to the first order conditions for maximum
likelihood estimation. Although they are formally different, the logit and
probit models are quite similar. We therefore expect the first three moments
to be the most informative about $\theta _{0}$. Moreover, we expect the
first moment to be the most important for determining $\widehat{\beta }_{0}$%
, and the second and third for determining $\widehat{\beta }_{1}$ and $%
\widehat{\beta }_{2}$, respectively.

Table \ref{Probit - optimal} shows results using the optimal weighting
matrix and Table \ref{Probit - diagonal} shows results using the diagonal
weighting matrix with the inverse of the moment variances on the diagonal.\footnote{%
We illustrate the proposed sensitivity measures through Monte Carlo
simulation of the expected values using $10^{7}$ simulated observations.}
We think of the latter as a practical alternative to the efficient weighting
matrix.

It is clear from Table \ref{Probit - optimal} that the first three moments
are indeed the most informative about $\beta _{0}$, $\beta _{1}$ and $\beta
_{2}$, respectively. As mentioned, this is expected since these moments
would be the first order conditions for maximum likelihood estimation of a
logit model.

The elements in the last three columns of $M_{1}$ in Table \ref{Probit -
optimal} are much smaller than the elements in the first three. This
suggests that the optimal GMM\ estimator is much less sensitive to
misspecification of the last three moments than to misspecification of the
first three moments. The reason is that the first three moments get almost
all the weight (in the corresponding logit model, they would literally get
all the weight). As expected, this is less pronounced in Table \ref{Probit -
diagonal}. The values of $\mathcal{E}_{2}$ in Tables \ref{Probit - optimal}
and \ref{Probit - diagonal} confirm that the efficient GMM estimator of $%
\theta _{0}$ is driven by the first three moments.\footnote{$\mathcal{E}_{2}$
in Tables \ref{Probit - optimal} and \ref{Probit - diagonal} differ only
because of simulation error.} Adding noise to the last three
moments has essentially no effect on the precision of the optimal GMM\
estimator of $\theta _{0}$, whereas adding noise to the first three elements
can have a big effect. The values of $\mathcal{E}_{3}$ in Table \ref{Probit
- diagonal} illustrate that the precision of the non-optimal GMM\ estimator
is less sensitive to noise in the last three moments (because they get
relatively less weight) and more sensitive to adding noise to the first
three moments (because they get relatively more weight).

Next, $\mathcal{E}_{4}$ and $\mathcal{E}_{5}$ suggest that leaving out, for
example, the second moment would increase the asymptotic variance of both
the efficient and the inefficient GMM estimator of $\beta _{1}$ by around
400 percent. This confirms that $E[ex_{1}]$ is instrumental for precise
estimation of $\beta _{1}$.

The final measure, $\mathcal{E}_{6}$ in Table \ref{Probit - optimal} is 0 by
construction. Since we are using the weighting matrix that minimizes the
variance of the estimator of each element of $\theta $, the derivative of
the variance with respect to the elements of the weighting matrix must be 0.
$\mathcal{E}_{6}$ in Table \ref{Probit - diagonal} shows that in this case,
the diagonal weighting matrix with the inverse of the moment variances on
the diagonal puts too little weight on the first three moments.

\subsection{Example 2: Duration Model}

The probit example in Section \ref{Probit} was chosen because it is an
example where we have good prior intuition about which moments matter for
what parameter. We now turn to an example where this is much less obvious.

Consider a duration, $T$, which follows a mixed proportional hazard model
with time-varying covariates and a Weibull as the baseline hazard
\begin{equation*}
h\left( t\right) =\alpha t^{\alpha -1}\exp \left( x^{\prime }\left( t\right)
\beta \right) \eta ,
\end{equation*}%
where $\alpha $ is the scale parameter which captures duration dependence and $%
x^{\prime }\left( t\right) \beta $ is the effect of the time-varying
explanatory variables. An example of a two-dimensional time-varying set of
explanatory variables could be
\begin{equation*}
x(t)=%
\begin{cases}
\begin{array}{ll}
(x_{1,1},x_{2,1}) & \text{if $t<t_{1}$} \\
(x_{1,2},x_{2,2}) & \text{if $t_{1}\leq t\leq t_{2}$} \\
\vdots & \vdots \\
(x_{1,k},x_{2,k}) & \text{if $t_{k-1}\leq t$}.%
\end{array}%
\end{cases}%
\end{equation*}
Finally, $\eta $ captures unobserved heterogeneity. Except for moment
assumptions, no assumptions are made on the distribution of $\eta$.

We then have the survival function for $T$,
\begin{equation*}
S\left( \left. t\right\vert x\left( \cdot \right) ,\eta \right) =\exp \left(
-\eta \int_{0}^{t}\alpha s^{\alpha -1}\exp \left( x^{\prime }\left( s\right)
\beta \right) ds\right) .
\end{equation*}%
Since%
\begin{equation*}
S\left( \left. T\right\vert x\left( \cdot \right) ,\eta \right) \sim U\left(
0,1\right) ,
\end{equation*}%
we have%
\begin{equation*}
\eta \int_{0}^{T}\alpha s^{\alpha -1}\exp \left( x^{\prime }\left( s\right)
\beta \right) ds\sim Exp\left( 1\right) \text{, conditional on }x\left(
\cdot \right) ,\eta
\end{equation*}%
or%
\begin{equation}
\log \left( \int_{0}^{T}\alpha s^{\alpha -1}\exp \left( x^{\prime }\left(
s\right) \beta \right) ds\right) \sim \log \left( Exp\left( 1\right) \right)
-\log \left( \eta \right) \text{, conditional on }x\left( \cdot \right)
,\eta .  \label{EQ: MPH-regression}
\end{equation}%
Here, $Exp\left( 1\right) $ denotes an exponentially distributed random
variable with mean 1, and $-\log \left( Exp\left( 1\right) \right) $ follows
a standard Gumbel distribution with $E\left[ -\log \left( Exp\left( 1\right)
\right) \right] =\gamma \thickapprox 0.57721$ (Euler's constant) and $V\left[
-\log \left( Exp\left( 1\right) \right) \right] =\pi ^{2}/6$.

Equation (\ref{EQ: MPH-regression}) suggests moment conditions of the type%
\begin{equation}
E\left[ \left( \log \left( \int_{0}^{T}\alpha s^{\alpha -1}\exp \left(
x^{\prime }\left( s\right) \beta \right) ds\right) +\gamma -\beta
_{0}\right) \psi \left( x\left( \cdot \right) \right) \right] =0
\label{EQ: Weibul Moments}
\end{equation}%
for functions of the covariates, $\psi $. Here, $\beta _{0}$ captures the
mean of $-\log \left( \eta \right) $ which is assumed to be finite.

When $x\left( t\right) $ is time-invariant, (\ref{EQ: MPH-regression})
becomes%
\begin{equation*}
\log \left( T^{\alpha }\exp \left( x^{\prime }\beta \right) \right) \sim
\log \left( Exp\left( 1\right) \right) -\log \left( \eta \right)
\end{equation*}%
or
\begin{equation*}
\log \left( T\right) =-x^{\prime }\left( \beta /\alpha \right) +\text{%
\textquotedblleft error\textquotedblright .}
\end{equation*}%
In other words, with time-invariant covariates the moments implied by (\ref%
{EQ: Weibul Moments}) do not identify $\left( \beta ,\alpha \right) $, but
only $\beta /\alpha $. It turns out that it is possible to estimate $\alpha $
by other methods (see, for example, \cite{honore1990}), but it is not
possible to estimate $\left( \beta ,\alpha \right) $ at the usual $\sqrt{n}$
rate (see \cite{Hahn1994}). This makes it interesting to investigate how
precision in estimation of $\left( \beta ,\alpha \right) $ depends on the
various moments in (\ref{EQ: Weibul Moments}) when $x$ does contain
time-varying covariates.

We consider a data generating process with one time-invariant and one
time-varying covariate. Specifically, $x\left( s\right) =\left( x_{1}\left(
s\right) ,x_{2}\left( s\right) \right) $ where
\begin{equation*}
x\left( s\right) =\left\{
\begin{array}{ccc}
\left( x_{1},x_{21}\right)  & \text{for} & s\leq 1 \\
\left( x_{1},x_{22}\right)  & \text{for} & 1<s\leq 2 \\
\left( x_{1},x_{23}\right)  & \text{for} & 2<s%
\end{array}%
\right.
\end{equation*}%
with $x_{1}=Z_{1}$, $x_{21}=Z_{2}$, $x_{22}=\left. \left(
x_{21}+Z_{3}\right) \right/ \sqrt{2}$ and $x_{23}=\left. \left(
x_{22}+Z_{4}\right) \right/ \sqrt{2}$.  $Z_{1}$ through $Z_{4}$ follow
standard normal distributions. The heterogeneity term, $\eta $, follows a
log-normal distribution, where the underlying normal has mean 0 and variance
$1/2$. $\eta $ is independent of $x\left( \cdot \right) $. Finally, $\beta
=\left( -1,1/\sqrt{2},1/\sqrt{2}\right) ^{\prime }$ and $\alpha =2$. With
this, the median duration is approximately 1.3, approximately 38\% of the
durations are less than 1, and 29\% greater than 2. This design is chosen
because it is a simple example with sizable unobserved heterogeneity and
duration dependence, and where we expect that the time-varying covariate
might have bite. The design is not meant to mimic any realistic empirical
example.

We again consider a moment-based estimator of $\theta =(\beta _{0}/\alpha
,\beta _{1}/\alpha ,\beta _{2}/\alpha ,\alpha )$ solving
\begin{equation*}
\hat{\theta}=\arg \min_{\theta }g(\theta )^{\prime }Wg(\theta ),
\end{equation*}%
where we use the five moments given by (\ref{EQ: Weibul Moments}) with $\psi
\left( x\left( \cdot \right) \right) =\left(
1,x_{1},x_{21},x_{22},x_{23}\right) $.

The sensitivity measures are given in Tables \ref{Weibul - optimal}\ and \ref%
{Weibul - diagonal}. In this design, the derivative of the first two moments
at the true parameter values are non-zero with respect to $\theta _{0\text{ }%
}$ and $\theta _{1}$, respectively. The derivatives are 0 with respect to
the other parameters. This implies that $G$ becomes singular when we exclude
either of the first two moments. This explains the extreme entries for $%
\mathcal{E}_{4}$ and $\mathcal{E}_{5}$ in Tables \ref{Weibul - optimal}\ and %
\ref{Weibul - diagonal}.

The conclusions from the remaining parts of the sensitivity measures are
fairly consistent. Most interestingly, the moments formed on the basis of
the time-varying covariates contribute to the identification of $\alpha$,
while the moment based on the time-invariant covariate does not. This is
exactly what the discussion above would predict. Interestingly, the first
moment is also important for $\alpha $. Presumably, this is because this
moment determines the estimate of the mean of the (log of the) unobserved
heterogeneity. It is well-known in the duration literature that unobserved
heterogeneity is poorly distinguished from duration dependence. As a result,
we do not consider this surprising.

\section{Application: Joint Retirement Planning\label{Sec: Application}}

In this section, we apply the proposed sensitivity measures to an extremely
simple structural model of the joint retirement planning of dual-earner
couples.

\subsection{Data and Institutional Setting}

We use the British Household Panel Survey (BHPS), which is a completed panel
of 18 waves collected from 1991 through 2009. In waves 11 and 16 of the BHPS,
each adult household member is asked, \emph{\textquotedblleft Even if this is
some time away, at what age do you expect you will retire?\textquotedblright
{}} We use this to measure the subjective retirement plans of each spouse.%
\footnote{%
The exact formulation in wave 11 is slightly different:\emph{%
\textquotedblleft At what age do you expect to retire/will you consider
yourself to be retired?\textquotedblright {}}} Based on the age at the
interview and the expected retirement age, we can calculate the expected
retirement year of each household member and use that to investigate joint
retirement plans.

Besides retirement plans, we use information in the BHPS on annual labor
market income, the number of visits to the general practitioner (GP),
subjective expectations about future health status, eligibility for an
employer provided pension scheme (EPP), and whether individuals save any of
their income in a private personal pension (PPP).\footnote{%
The EPP includes both defined and contributed benefit (DB and CB) plans and
we cannot distinguish between them. \cite{BlundellMeghirSmith2004} show,
however, that DB plans were most common in the U.K. in this period.}
Finally, we define individuals as highly skilled if they have completed the
first or second stage of tertiary education (ISCED codes 5 or 6).

We use information on households consisting of two opposite-sex household
members who are either married or cohabiting, and who meet the following
sample selection criteria: \emph{i)} Both members are between 40 and 59
years old when interviewed, \emph{ii)} At least one member is not retired at
the time of the interview, and \emph{iii) }Retirement plans are observed in
the age range 50 to 70 for at least one member not retired at the time of the
interview. If a household satisfies the criteria in both waves (11 and 16),
we use both survey responses in the analysis. We refer to each household
member as husband or wife, although we also include households, where
couples are cohabiting, but not necessarily married.

\subsubsection*{The State Pension Age (SPA)}

The state pension age (SPA) in the U.K. is the age where individuals become
eligible to receive state pension from the government. Individuals who have
reached SPA and contributed to the scheme for sufficiently many years are
eligible to receive a weekly transfer with no means testing. In 2009, the
weekly rate was around $\pounds 95$. See \cite{BozioCrawfordTetlow2010}, \cite%
{BlundellMeghirSmith2004} and \cite{CribbEmmersonTetlow2013} for excellent
descriptions of the pension system in the U.K.

The SPA was 65 for men and 60 for women until the implementation of the
Pension Act 1995. The Pension Act 1995 introduced an increase in the SPA of
women born after April 6, 1950. While the SPA for men was unaffected, the
SPA for women was gradually increased by one month every month (by date of
birth) until the SPA for women reached 65 for cohorts born later than
(including) 1955. See \cite{ThurleyKeen2017} for a comprehensive discussion
of the reform.\footnote{%
After the relevant waves in the BHPS (11 and 16) were conducted, the Pension
Act 2007 further increased the SPA for both men and women. Since the
respondents were interviewed before this reform was passed (most interviews
was done no later than 2006), we abstract from this and other subsequent
reforms.} Since this might affect individual expectations, our modelling
framework explicitly allows for an effect of the Pension Act 1995 on
retirement planning.

\subsubsection*{Descriptive Statistics}

\begin{figure}[!h]
\caption{Joint Retirement Planning}
\label{fig:JointPlanning}\centering
\includegraphics[width=0.47\textwidth]{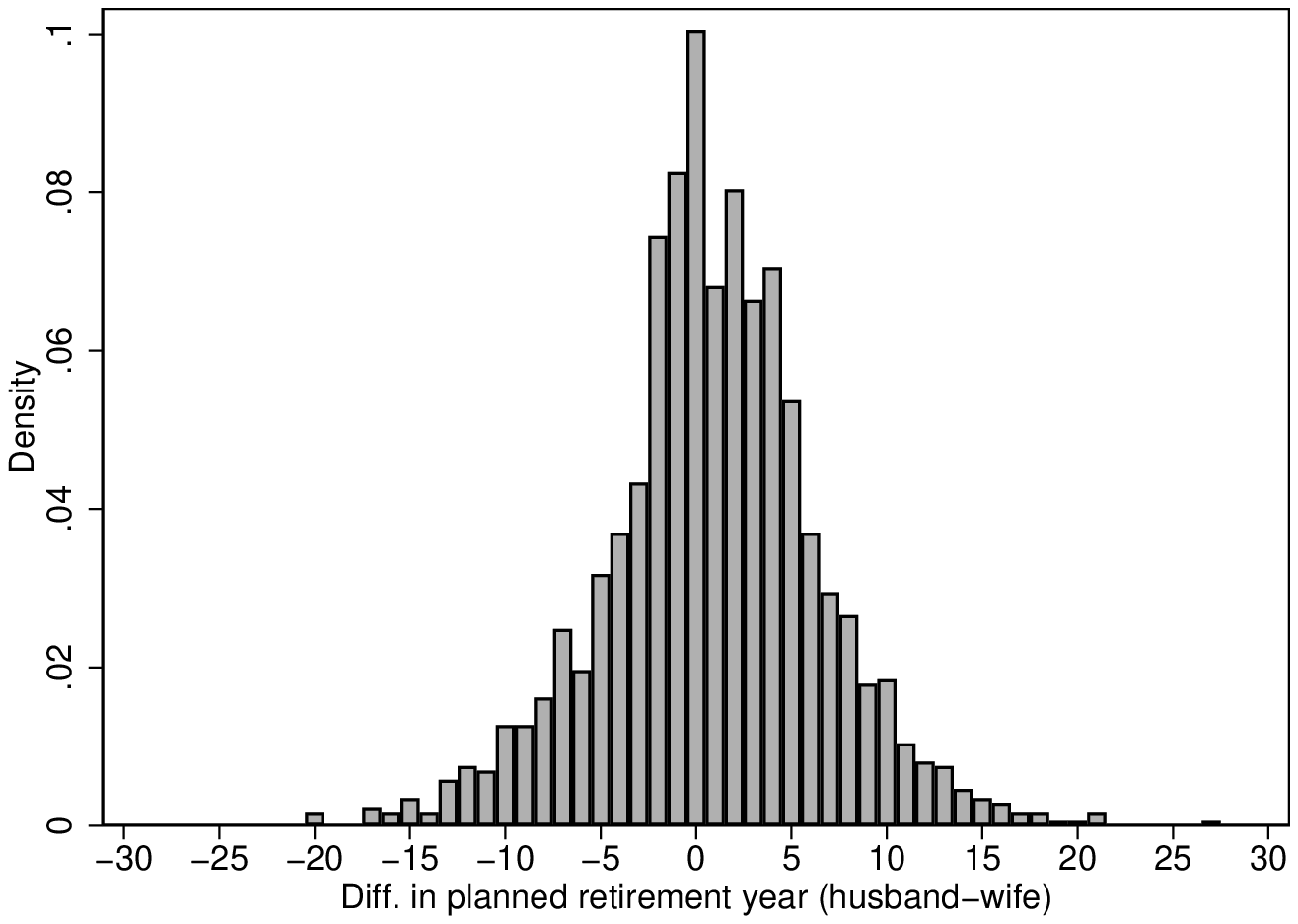}%
\includegraphics[width=0.47\textwidth]{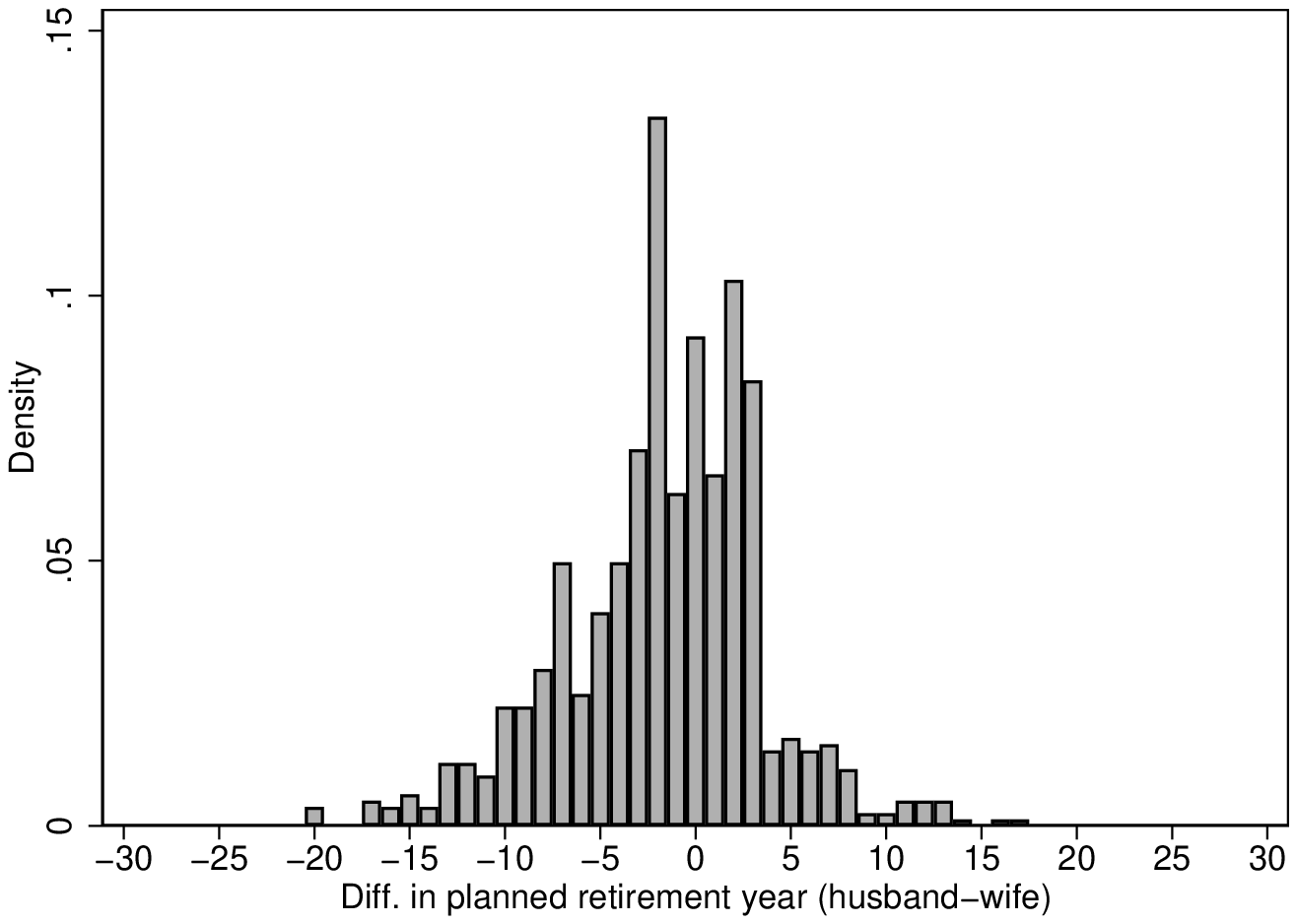}
\par
{\footnotesize {}{}}\noindent%
\begin{minipage}[t]{1\columnwidth}%
\emph{\footnotesize{}{}Notes: }{\footnotesize{}{}Figure \ref{fig:JointPlanning}
illustrates the difference in the }\emph{\footnotesize{}{}year}{\footnotesize{}{}
of retirement between husband and wife. The peak around zero indicates
joint retirement planning. Because the SPA of women is lower from
that of men for most cohorts, it is expected that the distribution
is right-tailed. The left panel illustrates the unconditional distribution and the right panel illustrates the distribution conditional on the husband being at least 2 years older than his spouse.}%
\end{minipage}
\end{figure}

Table \ref{tab:Descriptives} reports the descriptive statistics for the
variables that we use. All statistics are based on households in which both
members are not retired at the time of the interview, which is around 97
percent of our sample. Husbands in the estimation sample are approximately
1.5 years older than their wives, plan to retire two years later than their
wives (at age 63 on average), and the average difference in the planned
retirement year is approximately 0.83 years. This difference should be
viewed in light of the fact that the SPA of men is 65,
while it is substantially lower for most women in our sample and as low as
60 for women born before 1950. To illustrate simultaneous retirement
planning, Figure \ref{fig:JointPlanning} shows the distribution of the
difference in the planned year of retirement between husband and wife. The
left panel illustrates the unconditional distribution and the right panel
conditions on the husband being at least 2 years older than his wife. The
peak around zero indicates joint retirement planning, and the mass to the
right of zero likely stems from men being older than women and women having
a lower SPA. When conditioning on the husband being at least 2 years
older than his wife in the right panel, we see a substantial mass at 0 (same
planned retirement year); we now also see a substantial mass at $-2$ (same
planned retirement age).

Table \ref{tab:Descriptives} also shows that around 16 and 14 percent of men
and women, respectively, are classified as highly skilled, and we see  that
men tend to visit the GP much less than women. Interestingly, however, men
are more likely to expect their health to worsen in the future. The labor
income of husbands is around $\pounds 25,000$ while that of the wives is on average
around $\pounds 14,000$. Only around 13 percent of wives and 28 percent of husbands
contribute to a private pension (PPP), while around 47 percent of wives and
51 percent of husbands are eligible to some occupational retirement scheme
(EPP).

\subsection{A Model of Retirement Planning of Dual-Earner Households}

In this section, we formulate a discrete time version of the continuous time
bivariate duration model proposed in \cite{HonorePaula2018}. Specifically,
we parameterize the difference in the utility flow between being retired and
working. Utility maximization then gives an estimatable model for joint
retirement planning of couples.

\begin{figure}[!h]
\caption{Model Fit, Individual Retirement}
\label{fig:Model-Fit.}\includegraphics[width=0.47\textwidth]{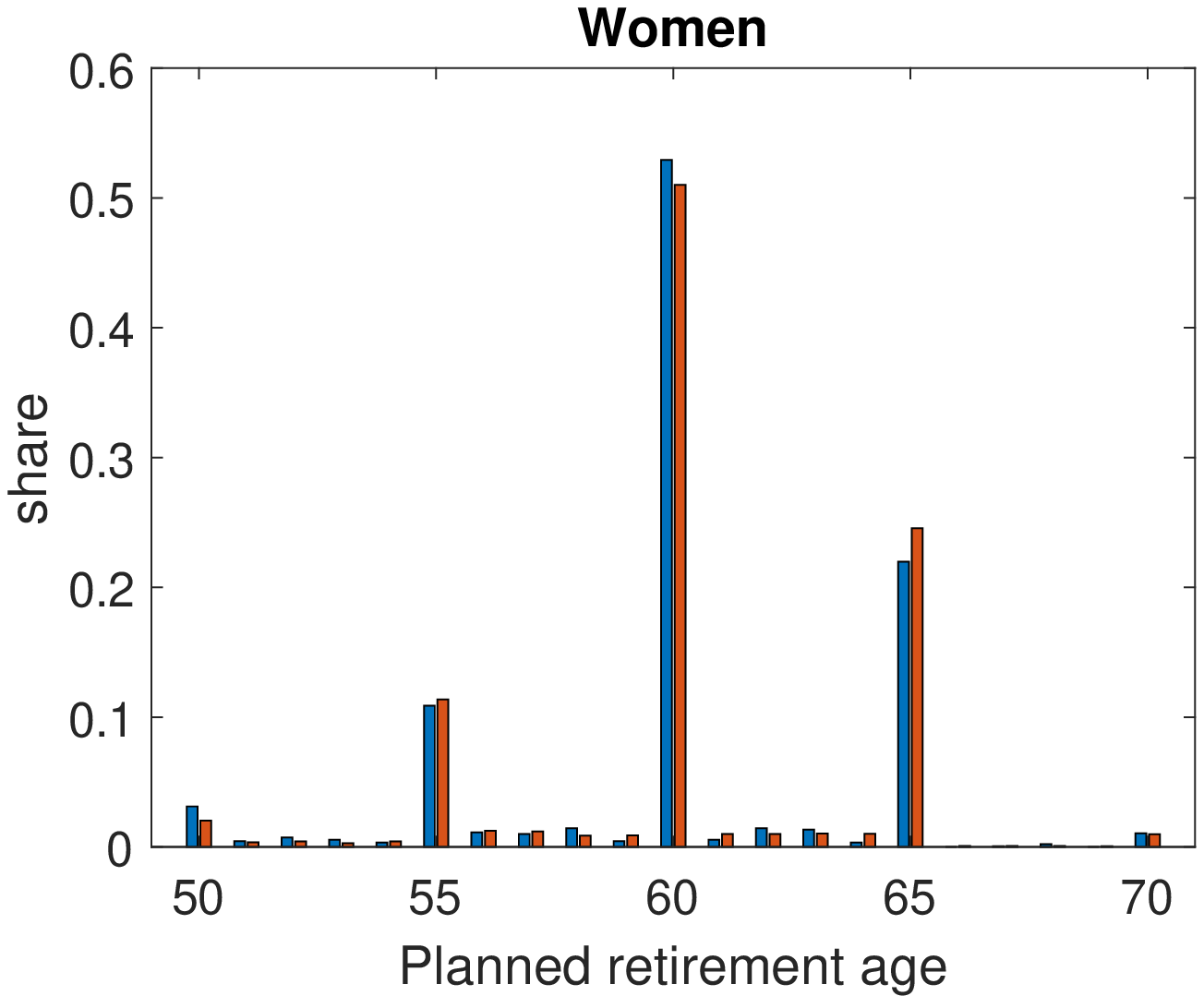}%
\includegraphics[width=0.47\textwidth]{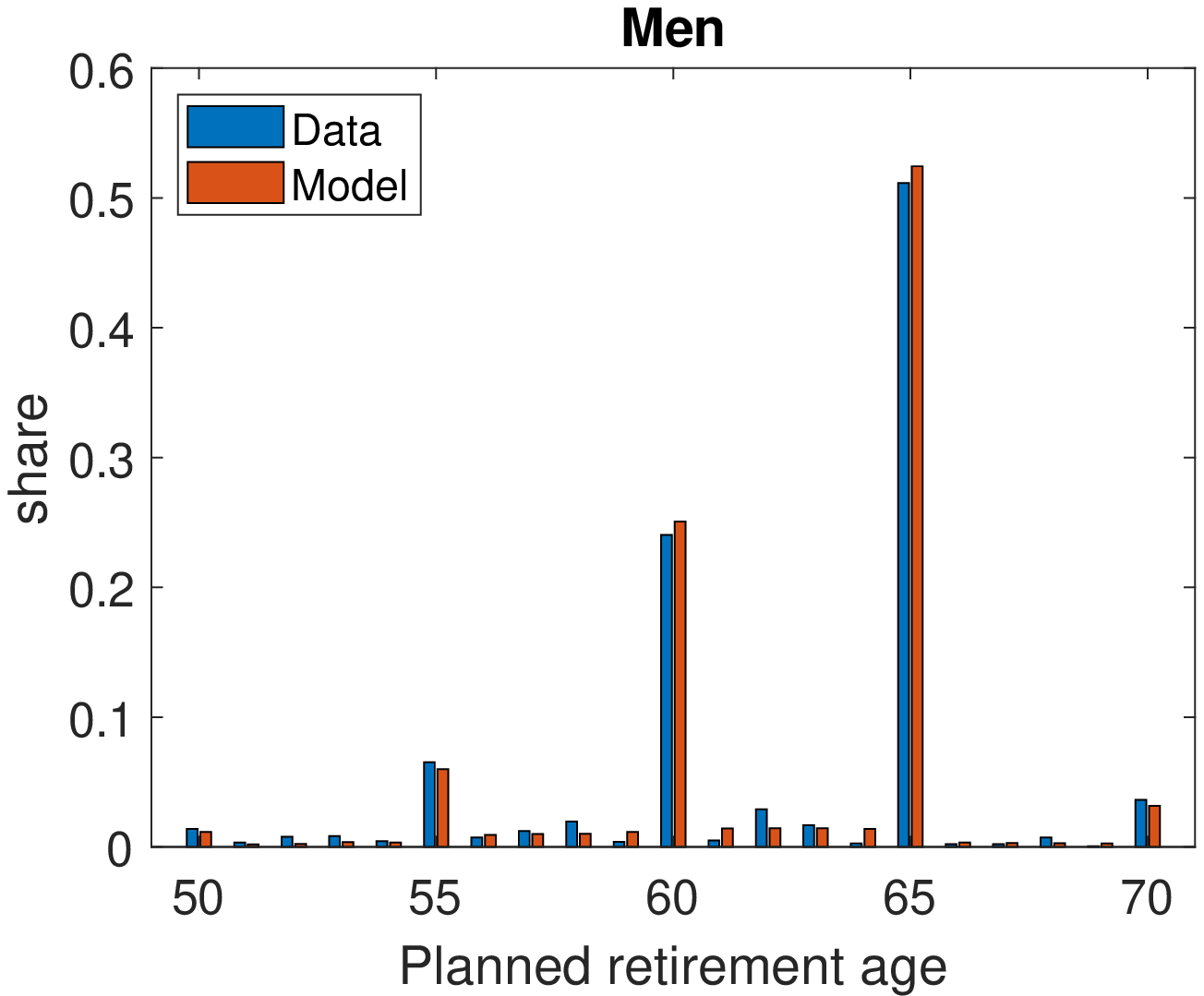}
\end{figure}

Consider first the husbands. We specify the difference in utility from being
retired in period $t$ compared to working as
\begin{equation*}
U_{h}(t,t_{w})=x_{h}^{\prime}\beta_{h}+\delta_{h}(t)+\gamma\mathbf{1}_{\{%
\mathcal{C}_{h}(t)\geq\mathcal{C}_{w}(t_{w})\}}+\varepsilon_{h},
\end{equation*}
where $\mathcal{C}_{h}\left(t\right)$ is the calendar time, $t_{w}$ is the
retirement age of the wife, and $\mathcal{C}_{w}\left(t_{w}\right)$ thus is
the calendar time at which the wife plans to retire. We interpret the term $%
\gamma\mathbf{1}_{\{\mathcal{C}(t)\geq\mathcal{C}(t_{w})\}}$ as a utility
externality that allows the husband to enjoy a higher utility flow from
planned retirement if the wife also plans to be retired at that time. We
parameterize the planned retirement age function, $\delta_{h}(t)$, as a
linear trend plus indicator functions for $t\geq55$, $t\geq60$ and $t\geq65$%
. The histograms in Figure \ref{fig:Model-Fit.} below suggest that these are
empirically important. We interpret the first two as reflecting either
social norms or heaping, while the third also reflects the fact that the SPA
for men is 65.

Similarly, the difference in utility flow for the wife is
\begin{equation*}
U_{w}(t,t_{h})=x_{w}^{\prime }\beta _{w}+\delta _{w}(t)+\gamma \mathbf{1}_{\{%
\mathcal{C}_{w}(t)\geq \mathcal{C}_{h}(t_{h})\}}+\alpha \mathbf{1}%
_{\{t_{w}\geq SPA_{w}\}}+\varepsilon _{w}.
\end{equation*}%
We again parameterize the function $\delta _{w}(t)$ as a linear trend plus
indicator functions for $t\geq 55$, $t\geq 60$ and $t\geq 65$. The term $%
\alpha \mathbf{1}_{\{t_{w}\geq SPA_{w}\}}$ reflects the idea that for women,
there is variation in the SPA as discussed above. This allows one to infer
the effect of the SPA separately from the dummies that reflect either
heaping or institutional features (e.g., early and statutory retirement ages) at 55, 60 and 65.

To close the model, we assume that $(\varepsilon _{h},\varepsilon _{w})$ is
jointly normal with mean zero and covariance matrix $\Omega $, where the
off-diagonal element of $\Omega $ captures possibly correlated retirement
preferences within households. We also assume that retirement is an
absorbing state. When the difference in utility from retirement compared to
working is increasing in age, this is not a binding constraint in the sense
that individuals would not want to re-enter the labor market once retired.

If a husband and a wife plan to retire at ages $r_{h}$ and $r_{w}$, their
discounted individual utilities are
\begin{equation*}
V_{h}(r_{h},r_{w})=\sum_{t=r_{h}}^{T_{max}}\rho ^{t-age_{h}}U_{h}(t,r_{w})
\end{equation*}%
for a husband aged $age_{h}$ and
\begin{equation*}
V_{w}(r_{w},r_{h})=\sum_{t=r_{w}}^{T_{max}}\rho ^{t-age_{w}}U_{w}(t,r_{h}),
\end{equation*}%
for a wife aged $age_{w}$. Finally, the optimal retirement plan for a
household is determined jointly as%
\begin{equation*}
(R_{h},R_{w})=\arg \max_{r_{h},r_{w}}\mathcal{A}%
(V_{h}(r_{h},r_{w}),V_{w}(r_{w},r_{h})),
\end{equation*}%
where $\mathcal{A}(\cdot ,\cdot )$ is a household aggregator. For the
estimation, we choose $\mathcal{A}(V_{h},V_{w})=V_{h}+\lambda V_{w}$ as in
the Nash bargaining setting from \cite{HonorePaula2018} or, more generally,
the collective model framework surveyed in \cite{BrowningChiapporiWeiss2014}.

It is clear that two scale normalizations are necessary in order to estimate
the model. First, the scale of $\mathcal{A}$ cannot be identified and we
therefore normalize the variance of $\varepsilon _{h}$ to be $\sigma
_{h}^{2}=1$. Secondly, the only effect of $\lambda $ is to re-scale all the
parameters in $V_{w}$. We therefore normalize $\lambda =1$. The model is
thus in effect unitary.

Our parameterization is inspired by the ordered probit model. Consider the
husbands. If $\gamma =0$ (such that there is no utility externality) and $%
\delta _{h}$ is increasing, then the utility maximation will lead to planned
retirement the first time $x_{h}^{\prime }\beta _{h}+\delta
_{h}(t)+\varepsilon _{h}>0$. In other words, the chosen planned retirement
age satisfies
\begin{equation*}
-\delta _{h}(R_{h})<x_{h}^{\prime }\beta _{h}+\varepsilon _{h}\leq -\delta
_{h}(R_{h}-1),
\end{equation*}%
which is exactly the ordered probit model. In that sense, the proposed model
is a generalization of the ordered probit model to a bivariate case with
simultaneity between the two outcomes.

\subsection{Indirect Inference Estimation}

We estimate the model's parameter vector $\theta =(\gamma ,\alpha ,\beta
_{h},\beta _{w},\delta _{h},\delta _{w},\sigma _{w}^{2},\sigma _{hw})$
through indirect inference\footnote{%
See, for example, \cite{smith1993}, and \cite{GourierouxMonfortRenault1993}.
While we use the Wald criterion function, indirect inference can also be
performed using other metrics (for example, the likelihood ratio or Lagrange
multiplier). See \cite{Smith2008}.},
\begin{equation*}
\hat{\theta}=\arg \min_{\theta \in \Theta }g(\theta )^{\prime }Wg(\theta ).
\end{equation*}%
The weighting matrix, $W$, is diagonal with the inverse of the variances of
the moments in the diagonal. $g(\theta )$ is a $K\times 1$ vector of
differences between statistics/moments in the data and identical moments
based on simulated data.

For each couple $i$, we simulate synthetic retirement plans by drawing $%
S_{sim}$ vectors of taste shocks $\varepsilon_{i}=\{\varepsilon_{i,h}^{(s)},%
\varepsilon_{i,w}^{(s)}\}_{s=1}^{S_{sim}}$ from the joint normal
distribution and calculate the value of all combinations of retirement ages
\begin{equation*}
V_{i}^{(s)}(r_{h},r_{w})=V_{h}(r_{h},r_{w}|x_{i},\varepsilon_{i,h}^{(s)},%
\varepsilon_{i,w}^{(s)})+\lambda
V_{w}(r_{w},r_{h}|x_{i},\varepsilon_{i,h}^{(s)},\varepsilon_{i,w}^{(s)}),
\end{equation*}
where the individual values are calculated as in (\ref{eq:Vh}) and (\ref%
{eq:Vw}). We then find the simulated retirement ages that maximize utility,
\begin{equation*}
(R_{i,h}^{(s)}(\theta),R_{i,w}^{(s)}(\theta))=\arg%
\max_{r_{h,}r_{w}}V_{i}^{(s)}(r_{h},r_{w})
\end{equation*}
for a given value of $\theta$.

To estimate the model parameters, we use four sets of auxiliary
models/moments with a total of $K=52$ elements in $g(\theta)$. We describe
in detail the construction of these moments in the supplemental material and
only list them here:

\begin{enumerate}
\item OLS coefficients from individual regressions of the planned retirement
age on own and spousal covariates $x_{i,h}$ and $x_{i,w}$ together with
indicators for the wife's birth cohort $\mathbf{1}\{1950<cohort_{w,i}%
\leq1954\}$ and $\mathbf{1}\{1955\leq cohort_{w,i}\}$.

\item The share of individuals planning to retire at ages 50-54, 55, 56-59,
60, 61-64, and 65, split by gender.

\item The covariance matrix of residuals from the regression in bullet 1
above for each household member.

\item The share of couples with retirement plans such that \emph{i)} the
wife plans to retire 1\textendash 2 years before her husband, \emph{ii)} the
husband plans to retire 1\textendash 2 years before his wife, or \emph{iii)}
the couple plan to retire in the same year.
\end{enumerate}

The first set of moments are primarily included to help estimate $\beta_{h}$%
, $\beta_{w}$, and $\alpha$ in the utility function. The second set of
moments are included primarily to help estimate the linear age trend and age
dummies in $\delta_{h}$ and $\delta_{w}$. The third set of moments are
primarily included to estimate the covariance of the preference shocks for
husband and wife, $\Omega$. Recall that we normalize $\sigma_{h}^{2}=1$ and
the remaining parameters in $\Omega$ are thus $\sigma_{w}^{2}$ and $%
\sigma_{hw}$. The final set of moments are included to estimate the value of
joint leisure, $\gamma$. We will use our proposed sensitivity measures below
to investigate these claims in a more systematic way.

\subsection{Empirical Results}

We use the BHPS data discussed above to estimate the model of joint
retirement planning of couples. We use the same moments as above and
simulate $S_{sim}=2000$ draws when approximating the expected moments. Table %
\ref{tab:Estimation-Results} reports the estimation results. We find a
positive value of coordination of around $\gamma \approx 0.026$, around two
to four times as large as the marginal utility from additional labor income
of $\pounds 1,000$ and significant at the $5\%$ level (\emph{p}-value of $0.02$).

Overall, the remaining statistically significant parameter estimates have
the expected signs. High skilled individuals value retirement less. Less
healthy people value retirement more, and having some form of pension
savings increase the value of retirement. Having an employer provided
pension (EPS) especially increases the utility from retirement compared to
working for husbands. Perhaps surprisingly, we find that higher earning women value
retirement more but this could proxy for higher wealth, which could lead to
a higher propensity to retire. All spousal variables seem to matter less and
are not statistically significant at most common significance levels.
Interestingly, we estimate a small positive and insignificant increase in
the expected retirement age of women in response to an increased SPA. This
goes in line with other studies finding a relatively low degree of awareness
of the reform (\cite{CrawfordTetlow2010}).

\begin{figure}[!ht]
\caption{Model Fit, Joint Retirement}
\label{fig:Model-Fit. joint}\centering{}\includegraphics[width=0.47%
\textwidth]{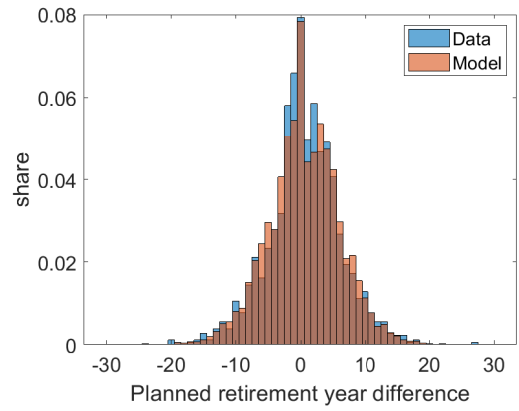}
\end{figure}

Figure \ref{fig:Model-Fit.} shows the histogram of planned retirement ages
for women and men. We see that the model does a quite good job fitting the
empirical distribution. Likewise, Figure \ref{fig:Model-Fit. joint} shows
the empirical and predicted distribution of retirement year differences
between couples. The predicted distribution matches the empirical one well,
although there are small deviations.

Table \ref{tab:Sensitivity-1} show the proposed sensitivity measures
together with the one proposed by \cite%
{AndrewsGentzkowShapiro2017_sensitivity}. We only report the measures for
the parameter of interest here: The value of joint leisure, $\gamma$. All
reported measures are scaled as discussed in Section \ref{Sec: Sensitivity
Measures}. The measure proposed by \cite%
{AndrewsGentzkowShapiro2017_sensitivity} is scaled such that $\mathcal{E}%
_{1}^{(j,k)}=M^{(j,k)}_1\sqrt{S^{(k,k)}}$.

Clearly, the moments which $\gamma$ is most sensitive to are related to
simultaneous retirement. In particular, we see from $\mathcal{E}_{4}$ and $%
\mathcal{E}_{5}$ that leaving out the moment ``the share planning to retire
the same year'' (moment 52) when estimating the model would increase the
asymptotic variance of $\gamma$ by a factor of 8. This confirms the intuition
that this moment is extremely informative about the value of joint leisure.
The share retiring within 2 years difference also seems important. In
particular, the correlation between the OLS regression residuals are
important. This is also intuitive since this moment captures a combination
of correlated shocks and preferences for joint leisure.

\section{Concluding Remarks\label{Sec: Conclusion}}

Structural econometric models are often estimated by matching moments that
depend on the parameters and on the data in a highly nonlinear way. This can
make it difficult to develop intuition for which moments of the data are
informative about which parameter. In this paper, we have proposed a number
of very simple sensitivity measures that are meant to shed light on this.

We have illustrated our measures in two artificial examples. The first is a
simple probit model and the second a mixed proportional hazard model with
time-varying covariates. The first illustrates that the proposed measures
are reasonable in a setting where the answer is rather obvious ex ante. The
second is chosen because it illustrates how the measures can be used to gain
insights, which are not so obvious.

We also illustrated the measures in a simple structural econometric model of
household retirement planning. This application is of independent interest
because it highlights the importance of modelling wives' and husbands'
retirement decisions jointly.

The econometric model for retirement that we develop can be interpreted as a
bivariate ordered choice model with simultaneity. Specifically, if the
\textquotedblleft utility externality\textquotedblright\ parameter is 0,
then the model that we estimate simplifies to a bivariate ordered probit
model. This may make it tractable in other applications.

\bibliographystyle{econometrica}
\bibliography{Refs_Joint}

\newpage

\begin{table}[!ht]
\begin{center}
\begin{threeparttable}
\caption{Sensitivity Measures, Probit Model, Optimal Weighting\label{Probit - optimal}}
\begin{tabular}{lrrrrrr} \toprule
& \multicolumn{6}{c}{Moment} \\ \cmidrule(lr){2-7}
& $ \mathbb{E}\left[ e \right] $   & $\mathbb{E}\left[ e x_{1}\right] $   & $\mathbb{E}\left[ e x_{2}\right] $    & $\mathbb{E}\left[ e x_{1}^{2} \right] $   & $\mathbb{E}\left[ e x_{1}x_{2}\right] $   & $\mathbb{E}\left[ e x_{2}^{2}\right] $   \\ \midrule
\multicolumn{7}{c}{$M_1$} \\ \cmidrule(lr){2-7}
 $\beta_0$& $    4.261 $ & $    1.475 $ & $    1.469 $ & $    0.192 $ & $    0.378 $ & $    0.184 $ \\
 $\beta_1$& $    1.190 $ & $    6.570 $ & $   -1.286 $ & $    0.223 $ & $    0.141 $ & $   -0.069 $ \\
 $\beta_1$& $    1.193 $ & $   -1.286 $ & $    6.567 $ & $   -0.073 $ & $    0.152 $ & $    0.214 $ \\
\cmidrule(lr){2-7} \multicolumn{7}{c}{$\mathcal{E}_2$} \\ \cmidrule(lr){2-7}
$\beta_0$& $    1.104 $ & $    0.088 $ & $    0.087 $ & $    0.003 $ & $    0.004 $ & $    0.003 $ \\
 $\beta_1$& $    0.060 $ & $    1.207 $ & $    0.046 $ & $    0.003 $ & $    0.000 $ & $    0.000 $ \\
 $\beta_1$& $    0.060 $ & $    0.046 $ & $    1.205 $ & $    0.000 $ & $    0.000 $ & $    0.003 $ \\
\cmidrule(lr){2-7} \multicolumn{7}{c}{$\mathcal{E}_3$} \\ \cmidrule(lr){2-7}
 $\beta_0$& $    1.104 $ & $    0.088 $ & $    0.087 $ & $    0.003 $ & $    0.004 $ & $    0.003 $ \\
 $\beta_1$& $    0.060 $ & $    1.207 $ & $    0.046 $ & $    0.003 $ & $    0.000 $ & $    0.000 $ \\
 $\beta_1$& $    0.060 $ & $    0.046 $ & $    1.205 $ & $    0.000 $ & $    0.000 $ & $    0.003 $ \\
\cmidrule(lr){2-7} \multicolumn{7}{c}{$\mathcal{E}_4$} \\ \cmidrule(lr){2-7}
 $\beta_0$& $    1.206 $ & $    0.292 $ & $    0.291 $ & $    0.005 $ & $    0.003 $ & $    0.005 $ \\
 $\beta_1$& $    0.065 $ & $    4.014 $ & $    0.155 $ & $    0.004 $ & $    0.006 $ & $    0.000 $ \\
 $\beta_1$& $    0.065 $ & $    0.153 $ & $    4.034 $ & $    0.000 $ & $    0.006 $ & $    0.004 $ \\
\cmidrule(lr){2-7} \multicolumn{7}{c}{$\mathcal{E}_5$} \\ \cmidrule(lr){2-7}
 $\beta_0$& $    1.203 $ & $    0.292 $ & $    0.291 $ & $    0.001 $ & $    0.003 $ & $    0.001 $ \\
 $\beta_1$& $    0.065 $ & $    4.014 $ & $    0.155 $ & $    0.001 $ & $    0.000 $ & $    0.000 $ \\
 $\beta_1$& $    0.065 $ & $    0.153 $ & $    4.034 $ & $    0.000 $ & $    0.000 $ & $    0.001 $ \\
 \cmidrule(lr){2-7} \multicolumn{7}{c}{$\mathcal{E}_6$} \\ \cmidrule(lr){2-7}
 $\beta_0$& $    0.000 $ & $   0.000 $ & $    0.000 $ & $    0.000 $ & $   0.000 $ & $    0.000 $ \\
 $\beta_1$& $   0.000 $ & $    0.000 $ & $    0.000 $ & $   0.000 $ & $   0.000 $ & $    0.000 $ \\
 $\beta_1$& $   0.000 $ & $    0.000 $ & $    0.000 $ & $    0.000 $ & $    0.000 $ & $    0.000 $ \\
 \bottomrule
\end{tabular}
\begin{footnotesize} \begin{tablenotes}
\item \emph{Notes: Simulations based on $10^7$ observations.}
\end{tablenotes} \end{footnotesize}
\end{threeparttable}
\end{center}
\end{table}

\newpage

\begin{table}[!ht]
\begin{center}
\begin{threeparttable}
\caption{Sensitivity Measures, Probit Model, Diagonal Weighting\label{Probit - diagonal}}
\begin{tabular}{lrrrrrr} \toprule
& \multicolumn{6}{c}{Moment} \\ \cmidrule(lr){2-7}
& $ \mathbb{E}\left[ e \right] $   & $\mathbb{E}\left[ e x_{1}\right] $   & $\mathbb{E}\left[ e x_{2}\right] $    & $\mathbb{E}\left[ e x_{1}^{2} \right] $   & $\mathbb{E}\left[ e x_{1}x_{2}\right] $   & $\mathbb{E}\left[ e x_{2}^{2}\right] $   \\ \midrule
\multicolumn{7}{c}{$M_1$} \\ \cmidrule(lr){2-7}
 $\beta_0$& $    3.374 $ & $    1.633 $ & $    1.630 $ & $    1.036 $ & $   -0.681 $ & $    1.035 $ \\
 $\beta_1$& $    1.354 $ & $    5.656 $ & $   -1.185 $ & $   -0.853 $ & $   -1.360 $ & $    0.882 $ \\
 $\beta_1$& $    1.351 $ & $   -1.185 $ & $    5.658 $ & $    0.881 $ & $   -1.360 $ & $   -0.851 $ \\
\cmidrule(lr){2-7} \multicolumn{7}{c}{$\mathcal{E}_2$} \\ \cmidrule(lr){2-7}
 $\beta_0$& $    1.104 $ & $    0.088 $ & $    0.087 $ & $    0.003 $ & $    0.004 $ & $    0.003 $ \\
 $\beta_1$& $    0.060 $ & $    1.207 $ & $    0.046 $ & $    0.003 $ & $    0.000 $ & $    0.000 $ \\
 $\beta_1$& $    0.060 $ & $    0.046 $ & $    1.205 $ & $    0.000 $ & $    0.000 $ & $    0.003 $ \\
\cmidrule(lr){2-7} \multicolumn{7}{c}{$\mathcal{E}_3$} \\ \cmidrule(lr){2-7}
 $\beta_0$& $    0.651 $ & $    0.101 $ & $    0.101 $ & $    0.080 $ & $    0.013 $ & $    0.080 $ \\
 $\beta_1$& $    0.071 $ & $    0.817 $ & $    0.036 $ & $    0.037 $ & $    0.034 $ & $    0.039 $ \\
 $\beta_1$& $    0.070 $ & $    0.036 $ & $    0.817 $ & $    0.039 $ & $    0.034 $ & $    0.037 $ \\
\cmidrule(lr){2-7} \multicolumn{7}{c}{$\mathcal{E}_4$} \\ \cmidrule(lr){2-7}
 $\beta_0$& $    1.076 $ & $    0.341 $ & $    0.340 $ & $   -0.010 $ & $   -0.011 $ & $   -0.011 $ \\
 $\beta_1$& $    0.042 $ & $    3.783 $ & $    0.116 $ & $   -0.038 $ & $   -0.031 $ & $   -0.028 $ \\
 $\beta_1$& $    0.042 $ & $    0.114 $ & $    3.802 $ & $   -0.028 $ & $   -0.032 $ & $   -0.038 $ \\
\cmidrule(lr){2-7} \multicolumn{7}{c}{$\mathcal{E}_5$} \\ \cmidrule(lr){2-7}
 $\beta_0$& $    1.203 $ & $    0.292 $ & $    0.291 $ & $    0.001 $ & $    0.003 $ & $    0.001 $ \\
 $\beta_1$& $    0.065 $ & $    4.014 $ & $    0.155 $ & $    0.001 $ & $    0.000 $ & $    0.000 $ \\
 $\beta_1$& $    0.065 $ & $    0.153 $ & $    4.034 $ & $    0.000 $ & $    0.000 $ & $    0.001 $ \\
 \cmidrule(lr){2-7} \multicolumn{7}{c}{$\mathcal{E}_6$} \\ \cmidrule(lr){2-7}
 $\beta_0$& $   -0.101 $ & $    0.002 $ & $    0.002 $ & $    0.040 $ & $    0.017 $ & $    0.041 $ \\
 $\beta_1$& $    0.011 $ & $   -0.142 $ & $    0.002 $ & $    0.044 $ & $    0.048 $ & $    0.037 $ \\
 $\beta_1$& $    0.011 $ & $    0.002 $ & $   -0.142 $ & $    0.037 $ & $    0.048 $ & $    0.044 $ \\
 \bottomrule
\end{tabular}
\begin{footnotesize} \begin{tablenotes}
\item \emph{Notes: Simulations based on $10^7$ observations.}
\end{tablenotes} \end{footnotesize}
\end{threeparttable}
\end{center}
\end{table}

\newpage

\begin{table}[!ht]
\begin{center}
\begin{threeparttable}
\caption{Sensitivity Measures, Weibull Model, Optimal Weighting\label{Weibul - optimal}}
\begin{tabular}{lrrrrr} \toprule
& \multicolumn{5}{c}{Moment} \\ \cmidrule(lr){2-6}
& ${E}\left[ e \right] $   & ${E}\left[ e x_{1}\right] $     & ${E}\left[ e x_{21}\right] $   & ${E}\left[ e x_{22}\right] $    & ${E}\left[ e x_{23}\right] $      \\ \midrule
& \multicolumn{5}{c}{$M_1$} \\ \cmidrule(lr){2-6}
$\beta_0$& $-0.503$& $0.001$& $3.375$& $-2.053$& $-1.934$\\
 $\beta_1$& $-0.000$& $-0.500$& $0.018$& $-0.015$& $-0.014$\\
 $\beta_2$& $-0.000$& $0.000$& $-0.228$& $-0.251$& $-0.184$\\
 $\alpha$& $-0.019$& $0.009$& $24.478$& $-15.092$& $-14.181$\\
\cmidrule(lr){2-6}
& \multicolumn{5}{c}{$\mathcal{E}_2$} \\ \cmidrule(lr){2-6}
 $\beta_0$& $0.028$& $0.000$& $1.282$& $0.474$& $0.421$\\
 $\beta_1$& $0.000$& $0.998$& $0.001$& $0.001$& $0.001$\\
 $\beta_2$& $0.000$& $0.000$& $0.155$& $0.187$& $0.100$\\
 $\alpha$& $0.000$& $0.000$& $1.299$& $0.494$& $0.436$\\
\cmidrule(lr){2-6}
& \multicolumn{5}{c}{$\mathcal{E}_3$} \\ \cmidrule(lr){2-6}
 $\beta_0$& $0.028$& $0.000$& $1.282$& $0.474$& $0.421$\\
 $\beta_1$& $0.000$& $0.998$& $0.001$& $0.001$& $0.001$\\
 $\beta_2$& $0.000$& $0.000$& $0.155$& $0.187$& $0.100$\\
 $\alpha$& $0.000$& $0.000$& $1.299$& $0.494$& $0.436$\\
\cmidrule(lr){2-6}
& \multicolumn{5}{c}{$\mathcal{E}_4$} \\ \cmidrule(lr){2-6}
 $\beta_0$& $> \negthinspace 100^*$& $> \negthinspace 100^*$& $4.841$& $0.196$& $0.274$\\
 $\beta_1$& $0.324^*$& $> \negthinspace 100^*$& $0.005$& $0.000$& $0.001$\\
 $\beta_2$& $0.012^*$& $> \negthinspace 100^*$& $0.584$& $0.077$& $0.065$\\
 $\alpha$& $3.935^*$& $> \negthinspace 100^*$& $4.904$& $0.203$& $0.284$\\
\cmidrule(lr){2-6}
& \multicolumn{5}{c}{$\mathcal{E}_5$} \\ \cmidrule(lr){2-6}
 $\beta_0$& $> \negthinspace 100^*$& $> \negthinspace 100^*$& $4.841$& $0.196$& $0.274$\\
 $\beta_1$& $0.324^*$& $> \negthinspace 100^*$& $0.005$& $0.000$& $0.001$\\
 $\beta_2$& $0.012^*$& $> \negthinspace 100^*$& $0.584$& $0.077$& $0.065$\\
 $\alpha$& $3.935^*$& $> \negthinspace 100^*$& $4.904$& $0.203$& $0.284$\\
\cmidrule(lr){2-6}
& \multicolumn{5}{c}{$\mathcal{E}_6$} \\ \cmidrule(lr){2-6}
 $\beta_0$& $0.000$& $-0.000$& $0.000$& $0.000$& $0.000$\\
 $\beta_1$& $-0.000$& $-0.000$& $0.000$& $0.000$& $0.000$\\
 $\beta_2$& $0.000$& $0.000$& $-0.000$& $0.000$& $0.000$\\
 $\alpha$& $-0.000$& $-0.000$& $0.000$& $0.000$& $0.000$\\
 \bottomrule
\end{tabular}
\begin{footnotesize} \begin{tablenotes}
\item \emph{Notes: Simulations based on $10^7$ observations. }
\item[$^*$] As mentioned in the text, large values of {$\mathcal{E}_4$} and {$\mathcal{E}_5$} suggest that the model is not identified after the moment has been removed from estimation.
\end{tablenotes} \end{footnotesize}
\end{threeparttable}
\end{center}
\end{table}

\newpage

\begin{table}[!ht]
\begin{center}
\begin{threeparttable}
\caption{Sensitivity Measures, Weibull Model, Diagonal Weighting\label{Weibul - diagonal}}
\begin{tabular}{lrrrrr} \toprule
& \multicolumn{5}{c}{Moment} \\ \cmidrule(lr){2-6}
& ${E}\left[ e \right] $   & ${E}\left[ e x_{1}\right] $     & ${E}\left[ e x_{21}\right] $   & ${E}\left[ e x_{22}\right] $    & ${E}\left[ e x_{23}\right] $      \\ \midrule
& \multicolumn{5}{c}{$M_1$} \\ \cmidrule(lr){2-6}
 $\beta_0$& $-0.503$& $0.001$& $3.066$& $-1.117$& $-2.679$\\
 $\beta_1$& $-0.000$& $-0.500$& $0.016$& $-0.010$& $-0.019$\\
 $\beta_2$& $-0.000$& $0.000$& $-0.234$& $-0.234$& $-0.197$\\
 $\alpha$& $-0.021$& $0.009$& $22.219$& $-8.255$& $-19.619$\\
\cmidrule(lr){2-6}
& \multicolumn{5}{c}{$\mathcal{E}_2$} \\ \cmidrule(lr){2-6}
 $\beta_0$& $0.028$& $0.000$& $1.282$& $0.474$& $0.421$\\
 $\beta_1$& $0.000$& $0.998$& $0.001$& $0.001$& $0.001$\\
 $\beta_2$& $0.000$& $0.000$& $0.155$& $0.187$& $0.100$\\
 $\alpha$& $0.000$& $0.000$& $1.299$& $0.494$& $0.436$\\
\cmidrule(lr){2-6}
& \multicolumn{5}{c}{$\mathcal{E}_3$} \\ \cmidrule(lr){2-6}
 $\beta_0$& $0.027$& $0.000$& $1.017$& $0.135$& $0.775$\\
 $\beta_1$& $0.000$& $0.998$& $0.001$& $0.000$& $0.001$\\
 $\beta_2$& $0.000$& $0.000$& $0.162$& $0.163$& $0.115$\\
 $\alpha$& $0.000$& $0.000$& $1.027$& $0.142$& $0.800$\\
\cmidrule(lr){2-6}
& \multicolumn{5}{c}{$\mathcal{E}_4$} \\ \cmidrule(lr){2-6}
 $\beta_0$& $> \negthinspace 100^*$& $> \negthinspace 100^*$& $4.612$& $0.149$& $0.224$\\
 $\beta_1$& $0.323^*$& $> \negthinspace 100^*$& $0.005$& $0.000$& $0.000$\\
 $\beta_2$& $0.011^*$& $> \negthinspace 100^*$& $0.583$& $0.077$& $0.065$\\
 $\alpha$& $3.737^*$& $> \negthinspace 100^*$& $4.667$& $0.155$& $0.232$\\
\cmidrule(lr){2-6}
& \multicolumn{5}{c}{$\mathcal{E}_5$} \\ \cmidrule(lr){2-6}
 $\beta_0$& $> \negthinspace 100^*$& $> \negthinspace 100^*$& $4.841$& $0.196$& $0.274$\\
 $\beta_1$& $0.324^*$& $> \negthinspace 100^*$& $0.005$& $0.000$& $0.001$\\
 $\beta_2$& $0.012^*$& $> \negthinspace 100^*$& $0.584$& $0.077$& $0.065$\\
 $\alpha$& $3.935^*$& $> \negthinspace 100^*$& $4.904$& $0.203$& $0.284$\\
\cmidrule(lr){2-6}
& \multicolumn{5}{c}{$\mathcal{E}_6$} \\ \cmidrule(lr){2-6}
 $\beta_0$& $0.000$& $-0.000$& $-0.049$& $-0.054$& $0.102$\\
 $\beta_1$& $0.000$& $0.000$& $-0.000$& $-0.000$& $0.000$\\
 $\beta_2$& $0.000$& $-0.000$& $0.002$& $-0.006$& $0.004$\\
 $\alpha$& $0.000$& $-0.000$& $-0.049$& $-0.056$& $0.105$\\
 \bottomrule
\end{tabular}
\begin{footnotesize} \begin{tablenotes}
\item \emph{Notes: Simulations based on $10^7$ observations.}
\item[$^*$] As mentioned in the text, large values of {$\mathcal{E}_4$} and {$\mathcal{E}_5$} suggest that the model is not identified after the moment has been removed from estimation.
\end{tablenotes} \end{footnotesize}
\end{threeparttable}
\end{center}
\end{table}

\newpage

\begin{table}[!ht]
\begin{center}
\begin{threeparttable}

\caption{Descriptive Statistics\label{tab:Descriptives}}

\begin{tabular}{lccccc}
\toprule
                    &        Mean&        Std.&         Min&         Max&        Obs.\\
\cmidrule(lr){2-6} Age, husband&      49.613&        5.53&          40&          59&        1730\\
Age, wife           &      48.128&        5.34&          40&          59&        1730\\
Planned retirement age, husband&      62.606&        3.87&          50&          70&        1730\\
Planned retirement age, wife&      60.301&        3.72&          50&          70&        1730\\
Diff. in planned retirement year (husband-wife)&       0.823&        5.71&         -20&          27&        1730\\
High skilled, husband&       0.157&        0.36&           0&           1&        1730\\
High skilled, wife  &       0.139&        0.35&           0&           1&        1730\\
10+ GP visits, husband&       0.039&        0.19&           0&           1&        1729\\
10+ GP visits, wife &       0.080&        0.27&           0&           1&        1729\\
Expect worse health, husband&       0.182&        0.39&           0&           1&        1641\\
Expect worse health, wife&       0.115&        0.32&           0&           1&        1645\\
Labor income (\pounds1,000), husband&      25.248&       17.12&           0&         244&        1600\\
Labor income (\pounds1,000), wife&      13.815&       10.78&           0&         109&        1442\\
Private pension, husband&       0.280&        0.45&           0&           1&        1730\\
Private pension, wife&       0.134&        0.34&           0&           1&        1730\\
Employer pension, husband&       0.514&        0.50&           0&           1&        1730\\
Employer pension, wife&       0.466&        0.50&           0&           1&        1730\\

\bottomrule
\end{tabular}

\begin{footnotesize} \end{footnotesize} \end{threeparttable}
\end{center}
\end{table}

\newpage 

\begin{table}[!ht]
\centering
\begin{threeparttable}\caption{Estimation Results, Indirect Inference\label{tab:Estimation-Results}}

\begin{tabular}{clrrrr} \toprule 
&& \multicolumn{2}{c}{Husband} & \multicolumn{2}{c}{Wife} \\ \hline 
$\gamma$ & Joint leisure & $0.026$ & ($0.011$) & $0.026$ & ($0.011$) \\ 
$\alpha$ & SPA age & \multicolumn{1}{c}{$-$} & \multicolumn{1}{c}{$-$} & $0.105$ & ($0.122$) \\ 
\multicolumn{6}{l}{} \\ 
\multicolumn{6}{l}{\emph{Explanatory variables ($\beta$)}} \\ 
\multicolumn{2}{l}{High skilled} & $-0.129$ & ($0.100$) & $-0.148$ & ($0.110$) \\ 
\multicolumn{2}{l}{10+ GP visits} & $0.315$ & ($0.291$) & $0.152$ & ($0.157$) \\ 
\multicolumn{2}{l}{Expect worse health} & $0.091$ & ($0.112$) & $0.001$ & ($0.109$) \\ 
\multicolumn{2}{l}{Labor income (1,000\pounds)} & $0.006$ & ($0.003$) & $0.011$ & ($0.005$) \\ 
\multicolumn{2}{l}{Has private pension (PPP)} & $0.194$ & ($0.092$) & $-0.005$ & ($0.084$) \\ 
\multicolumn{2}{l}{Has employer provided pension (EPS)} & $0.610$ & ($0.089$) & $-0.044$ & ($0.060$) \\ 
\multicolumn{2}{l}{Birth year (minus 1955)} & $0.005$ & ($0.005$) & $-0.005$ & ($0.007$) \\ 
\multicolumn{2}{l}{Labor income (1,000\pounds), spouse} & $0.005$ & ($0.004$) & $0.003$ & ($0.003$) \\ 
\multicolumn{2}{l}{Has private pension (PPP), spouse} & $0.074$ & ($0.093$) & $-0.005$ & ($0.077$) \\ 
\multicolumn{2}{l}{Has employer provided pension (EPS), spouse} & $0.171$ & ($0.076$) & $0.013$ & ($0.080$) \\ 
\multicolumn{6}{l}{} \\ 
\multicolumn{6}{l}{\emph{Age variables ($\delta$)}} \\ 
\multicolumn{2}{l}{Constant} & $-2.413$ & ($0.128$) & $-1.667$ & ($0.474$) \\ 
\multicolumn{2}{l}{Time trend (minus 25)} & $0.036$ & ($0.004$) & $0.020$ & ($0.007$) \\ 
\multicolumn{2}{l}{Retirement age 55 dummy} & $0.632$ & ($0.068$) & $0.729$ & ($0.177$) \\ 
\multicolumn{2}{l}{Retirement age 60 dummy} & $0.867$ & ($0.038$) & $1.323$ & ($0.362$) \\ 
\multicolumn{2}{l}{Retirement age 65 dummy} & $1.978$ & ($0.078$) & $1.452$ & ($0.418$) \\ 
\multicolumn{6}{l}{} \\ 
$\sigma$ & variance & $1.000$ && $0.917$ &\\ 
$\sigma_{hw}$ & covariance & $0.359$ & & $0.359$ & \\ 
\bottomrule \end{tabular}

\begin{footnotesize} \begin{tablenotes}

\item \emph{Notes:} The table reports the estimated simultaneous
retirement planning model using the BHPS data using indirect inference.
Asymptotic standard errors reported in brackets.

\end{tablenotes} \end{footnotesize} \end{threeparttable}
\end{table}

\newpage

\begin{table}[!ht]
\centering
\begin{threeparttable}\caption{Sensitivity of $\gamma$\label{tab:Sensitivity-1}}

\begin{tabular}{ll*{6}{r}} 
 \toprule \multicolumn{2}{l}{Moment} & \multicolumn{1}{c}{$\mathcal{E}_1$} & \multicolumn{1}{c}{$\mathcal{E}_2$} & \multicolumn{1}{c}{$\mathcal{E}_3$} & \multicolumn{1}{c}{$\mathcal{E}_4$} & \multicolumn{1}{c}{$\mathcal{E}_5$} & \multicolumn{1}{c}{$\mathcal{E}_6$} \\ \cmidrule(lr){3-8} 
\multicolumn{2}{l}{\emph{Regression, husband}} & \multicolumn{6}{c}{} \\ 
1 & Constant & $-0.006$ & $0.259$ & $0.000$ & $-0.001$ & $0.009$ & $0.001$ \\ 
2 & High skilled, husband & $0.075$ & $0.041$ & $0.024$ & $0.270$ & $0.216$ & $-0.010$ \\ 
3 & 10+ GP visits, husband & $0.018$ & $0.008$ & $0.001$ & $0.065$ & $0.049$ & $-0.002$ \\ 
4 & Expect worse health, husband & $-0.003$ & $0.001$ & $0.000$ & $-0.008$ & $0.009$ & $0.000$ \\ 
5 & Labor income, husband & $0.007$ & $0.000$ & $0.000$ & $-0.005$ & $0.000$ & $0.001$ \\ 
6 & Has private pension, husband & $-0.066$ & $0.023$ & $0.019$ & $0.256$ & $0.097$ & $-0.004$ \\ 
7 & Has employer provided pension, husband & $-0.032$ & $0.005$ & $0.004$ & $0.040$ & $0.003$ & $-0.005$ \\ 
8 & Birth year (minus 1955), husband & $-0.036$ & $0.001$ & $0.006$ & $-0.006$ & $0.001$ & $0.004$ \\ 
9 & High skilled, wife & $0.015$ & $0.001$ & $0.001$ & $-0.004$ & $0.001$ & $0.003$ \\ 
10 & 10+ GP visits, wife & $-0.007$ & $0.001$ & $0.000$ & $-0.002$ & $0.001$ & $0.002$ \\ 
11 & Expect worse health, wife & $-0.005$ & $0.005$ & $0.000$ & $0.001$ & $0.004$ & $-0.001$ \\ 
12 & Labor income, wife & $-0.017$ & $0.005$ & $0.001$ & $0.010$ & $0.009$ & $-0.001$ \\ 
13 & Has private pension, wife & $0.105$ & $0.039$ & $0.048$ & $0.904$ & $1.033$ & $0.014$ \\ 
14 & Has employer provided pension, wife & $0.009$ & $0.005$ & $0.000$ & $0.022$ & $0.011$ & $-0.001$ \\ 
15 & Birth year, wife & $0.003$ & $0.003$ & $0.000$ & $0.001$ & $0.000$ & $-0.001$ \\ 
16 & Birth year, wife in 1951--1955 & $-0.003$ & $0.072$ & $0.000$ & $-0.002$ & $0.030$ & $0.002$ \\ 
17 & Birth year, wife later than 1955 & $-0.005$ & $0.130$ & $0.000$ & $-0.001$ & $0.006$ & $0.001$ \\ 
\multicolumn{2}{l}{\emph{Regression, wife}} & \multicolumn{6}{c}{} \\ 
18 & Constant & $-0.024$ & $0.075$ & $0.002$ & $-0.006$ & $0.005$ & $0.005$ \\ 
19 & High skilled, husband & $0.025$ & $0.013$ & $0.003$ & $0.002$ & $0.007$ & $-0.000$ \\ 
20 & 10+ GP visits, husband & $0.011$ & $0.005$ & $0.001$ & $-0.002$ & $0.004$ & $0.002$ \\ 
21 & Expect worse health, husband & $-0.005$ & $0.007$ & $0.000$ & $0.000$ & $0.004$ & $-0.000$ \\ 
22 & Labor income, husband & $-0.012$ & $0.007$ & $0.001$ & $0.008$ & $0.015$ & $-0.001$ \\ 
23 & Has private pension, husband & $0.106$ & $0.044$ & $0.050$ & $0.899$ & $0.733$ & $0.012$ \\ 
24 & Has employer provided pension, husband & $-0.062$ & $0.012$ & $0.017$ & $0.003$ & $0.048$ & $0.009$ \\ 
25 & Birth year (minus 1955), husband & $0.023$ & $0.090$ & $0.002$ & $0.009$ & $0.032$ & $-0.005$ \\ 
26 & High skilled, wife & $0.064$ & $0.015$ & $0.018$ & $0.053$ & $0.054$ & $0.003$ \\ 
27 & 10+ GP visits, wife & $0.009$ & $0.001$ & $0.000$ & $0.033$ & $0.010$ & $-0.001$ \\ 
28 & Expect worse health, wife & $0.057$ & $0.016$ & $0.014$ & $0.276$ & $0.155$ & $0.001$ \\ 
29 & Labor income, wife & $-0.017$ & $0.007$ & $0.001$ & $0.028$ & $0.025$ & $-0.001$ \\ 
30 & Has private pension, wife & $-0.060$ & $0.014$ & $0.016$ & $0.755$ & $0.337$ & $0.003$ \\ 
31 & Has employer provided pension, wife & $0.009$ & $0.011$ & $0.000$ & $0.034$ & $0.045$ & $-0.001$ \\ 
32 & Birth year, wife & $-0.028$ & $0.062$ & $0.003$ & $0.011$ & $0.009$ & $-0.005$ \\ 
33 & Birth year, wife in 1951--1955 & $-0.013$ & $0.039$ & $0.001$ & $0.003$ & $0.013$ & $-0.002$ \\ 
34 & Birth year, wife later than 1955 & $-0.024$ & $0.021$ & $0.003$ & $-0.003$ & $0.001$ & $0.002$ \\ 
\bottomrule \end{tabular}

\begin{footnotesize} \begin{tablenotes}

\item \emph{Notes:} The table reports the sensitivity measures of
$\gamma$ for the estimated joint retirement planning model.

\end{tablenotes} \end{footnotesize} \end{threeparttable}
\end{table}

\newpage

\addtocounter{table}{-1}

\begin{table}[!ht]
\centering
\begin{threeparttable}\caption{Sensitivity of $\gamma$ (continued)\label{tab:Sensitivity-2}}

\begin{tabular}{ll*{6}{r}} 
 \toprule \multicolumn{2}{l}{Moment} & \multicolumn{1}{c}{$\mathcal{E}_1$} & \multicolumn{1}{c}{$\mathcal{E}_2$} & \multicolumn{1}{c}{$\mathcal{E}_3$} & \multicolumn{1}{c}{$\mathcal{E}_4$} & \multicolumn{1}{c}{$\mathcal{E}_5$} & \multicolumn{1}{c}{$\mathcal{E}_6$} \\ \cmidrule(lr){3-8} 
\multicolumn{2}{l}{\emph{Retirement age, husband}} & \multicolumn{6}{c}{} \\ 
35 & Share at ages 50--54 & $0.005$ & $0.000$ & $0.000$ & $0.003$ & $0.000$ & $-0.001$ \\ 
36 & Share at age 55 & $0.030$ & $0.000$ & $0.004$ & $0.036$ & $0.000$ & $-0.009$ \\ 
37 & Share at ages 56--59 & $-0.040$ & $0.081$ & $0.007$ & $0.050$ & $0.042$ & $-0.009$ \\ 
38 & Share at age 60 & $-0.003$ & $0.043$ & $0.000$ & $-0.001$ & $0.005$ & $0.000$ \\ 
39 & Share at ages 61--64 & $0.015$ & $0.000$ & $0.001$ & $0.032$ & $0.000$ & $-0.003$ \\ 
40 & Share at age 65 & $0.005$ & $0.027$ & $0.000$ & $0.002$ & $0.004$ & $-0.000$ \\ 
\multicolumn{2}{l}{\emph{Retirement age, wife}} & \multicolumn{6}{c}{} \\ 
41 & Share at ages 50--54 & $0.024$ & $0.011$ & $0.002$ & $0.007$ & $0.007$ & $-0.001$ \\ 
42 & Share at age 55 & $0.010$ & $0.067$ & $0.000$ & $0.007$ & $0.019$ & $-0.002$ \\ 
43 & Share at ages 56--59 & $-0.024$ & $0.014$ & $0.003$ & $-0.009$ & $0.007$ & $0.002$ \\ 
44 & Share at age 60 & $-0.001$ & $0.250$ & $0.000$ & $-0.001$ & $0.020$ & $0.000$ \\ 
45 & Share at ages 61--64 & $0.024$ & $0.040$ & $0.003$ & $0.058$ & $0.037$ & $-0.002$ \\ 
46 & Share at age 65 & $0.006$ & $0.082$ & $0.000$ & $0.004$ & $0.018$ & $-0.001$ \\ 
\multicolumn{2}{l}{\emph{Simultaneous retirement}} & \multicolumn{6}{c}{} \\ 
47 & var$(e_h)$ & $-0.008$ & $0.002$ & $0.000$ & $-0.004$ & $0.000$ & $0.001$ \\ 
48 & var$(e_w)$ & $-0.005$ & $0.076$ & $0.000$ & $-0.005$ & $0.021$ & $0.001$ \\ 
49 & cov$(e_h,e_w)$ & $-0.145$ & $0.204$ & $0.092$ & $0.757$ & $0.557$ & $-0.040$ \\ 
50 & diff [-2,-1] & $0.018$ & $0.035$ & $0.001$ & $0.018$ & $0.037$ & $-0.007$ \\ 
51 & diff [1,2] & $-0.113$ & $0.000$ & $0.056$ & $-0.067$ & $0.000$ & $0.077$ \\ 
52 & Joint retirement & $0.343$ & $0.684$ & $0.516$ & $8.019$ & $5.541$ & $-0.030$ \\ 
\bottomrule \end{tabular}

\begin{footnotesize} \begin{tablenotes}

\item \emph{Notes:} The table reports the sensitivity measures of
$\gamma$ for the estimated joint retirement planning model.

\end{tablenotes} \end{footnotesize} \end{threeparttable}
\end{table}

\FloatBarrier

\bigskip

\newpage

\newpage \FloatBarrier

\begin{center}
{\Huge {Online supplemental material} }
\end{center}

\section*{Definition of Moments used for Estimation\label{sec: Moments}}

\paragraph*{Individual OLS Moment Conditions.}

Let $R_{i,j}$ denote the planned retirement age of member $j$ in household $%
i $ and $X_{i}=(1,x_{i,h}^{\prime },x_{i,w}^{\prime },\mathbf{1}%
\{1950<cohort_{w,i}\leq 1954\},\mathbf{1}\{1955\leq cohort_{w,i}\})^{\prime
} $ denote the set of control variables. We include as the first set of
moments
\begin{equation*}
\mathcal{M}_{1}(\theta )=\frac{1}{N}\sum_{i=1}^{N}\frac{1}{S}%
\sum_{s=1}^{S}\left(
\begin{array}{c}
X_{i}e_{i,h}^{(s)}(\theta ) \\
X_{i}e_{i,w}^{(s)}(\theta )%
\end{array}%
\right)
\end{equation*}%
where, for $j=\{h,w\}$,
\begin{equation*}
e_{i,j}^{(s)}(\theta )=R_{i,j}^{(s)}(\theta )-X_{i}^{\prime }\hat{\beta}%
_{j}^{OLS}
\end{equation*}%
where $\hat{\beta}_{j}^{OLS}=(X^{\prime }X)^{-1}X^{\prime }R_{j}$ are the
OLS regression coefficients using the data.

\paragraph*{Covariance Matrix of Regression Residuals.}

The second set of moments are related to the regression above. Particularly,
we include as the second set of moments the simulated difference in the
moments of the error terms
\begin{equation*}
\mathcal{M}_{2}(\theta )=\frac{1}{N}\sum_{i=1}^{N}\frac{1}{S}%
\sum_{s=1}^{S}\left(
\begin{array}{c}
e_{i,h}^{2}-(e_{i,h}^{(s)}(\theta ))^{2} \\
e_{i,w}^{2}-(e_{i,w}^{(s)}(\theta ))^{2} \\
e_{i,h}e_{i,w}-e_{i,h}^{(s)}(\theta )e_{i,w}^{(s)}(\theta )%
\end{array}%
\right)
\end{equation*}%
where $e_{i,j}=R_{i,j}-X_{i}^{\prime }\hat{\beta}_{j}^{OLS}$ is the
residuals from the regression using the data.

\paragraph*{Planned Retirement Age Groups.}

Next, we include the share of individuals retiring in 6 particular
age-groups, $k=\{50-54,55,56-59,60,61-64,65\}$. Denote as $\mathcal{S}%
_{i,j}=(d_{i,j,1},\dots ,d_{i,j,6})^{\prime }$ the 6-element column vector
of dummies where $d_{i,j,k}$ is one if member $j$ in household $i$ is in
group $k$ and zero otherwise. Likewise, denote $\mathcal{S}%
_{i,j}^{(s)}(\theta )$ as the simulated counter-part of this set of dummies.
We then include as the third set of moments,
\begin{equation*}
\mathcal{M}_{3}(\theta )=\frac{1}{N}\sum_{i=1}^{N}\frac{1}{S}%
\sum_{s=1}^{S}\left(
\begin{array}{c}
\mathcal{S}_{i,h}-\mathcal{S}_{i,h}^{(s)}(\theta ) \\
\mathcal{S}_{i,w}-\mathcal{S}_{i,w}^{(s)}(\theta )%
\end{array}%
\right).
\end{equation*}

\paragraph*{Simultaneous retirement.}

The final moments included relate to the retirement timing of couples.
Defining the retirement calendar year as $\mathcal{C}_{i,m}$ and the
simulated counterpart as $\mathcal{C}_{i,m}^{(s)}(\theta)$, the final
moments are
\begin{equation*}
\mathcal{M}_{4}(\theta)=\frac{1}{N}\sum_{i=1}^{N}\frac{1}{S}%
\sum_{s=1}^{S}\left(%
\begin{array}{c}
\mathbf{1}\{\mathcal{C}_{i,h}-\mathcal{C}_{i,w}\in\{-2,-1\}\}-\mathbf{1}\{%
\mathcal{C}_{i,h}^{(s)}(\theta)-\mathcal{C}_{i,w}^{(s)}(\theta)\in\{-2,-1\}\}
\\
\mathbf{1}\{\mathcal{C}_{i,h}-\mathcal{C}_{i,w}\in\{1,2\}\}-\mathbf{1}\{%
\mathcal{C}_{i,h}^{(s)}(\theta)-\mathcal{C}_{i,w}^{(s)}(\theta)\in\{1,2\}\}
\\
\mathbf{1}\{\mathcal{C}_{i,h}=\mathcal{C}_{i,w}\}-\mathbf{1}\{\mathcal{C}%
_{i,h}^{(s)}(\theta)=\mathcal{C}_{i,w}^{(s)}(\theta)\}%
\end{array}%
\right).
\end{equation*}

Stacking all moments together gives
\begin{equation*}
g(\theta)=(\mathcal{M}_{1}(\theta),\mathcal{M}_{2}(\theta),\mathcal{M}%
_{3}(\theta),\mathcal{M}_{4}(\theta))^{\prime}
\end{equation*}
and the estimator of $\theta$ is
\begin{equation*}
\hat{\theta}=\arg\min_{\theta\in\Theta}g(\theta)^{\prime}Wg(\theta)
\end{equation*}
where we use as weighting a matrix, $W$, the inverse of the bootstrapped
variances of the moments on the diagonal and zero everywhere else.

We solve the minimization problem by successively applying different
minimization routines in Matlab. We perform the sequence of estimators four
times and report the estimates yielding the lowest criteria function. For
each of the four estimation runs, we start with MATLABs \texttt{particleswarm%
} which is a ``global'' optimization routine using randomization to search
through the parameter space. We use 80 particles and switch to Nelder-Mead (%
\texttt{fminsearch} in MATLAB) using the best candidates from the converged
particleswarm. We use $S_{sim}=100$ simulation draws for this estimation.
After the four sequences of these two algorithms, we increase the number of
simulation draws to $S_{sim}=2,000$ and do one final Nelder-Mead
minimization starting at the parameters yielding the lowest objective
function over the four sequences of estimators. We then report the parameter
values that solves this final minimization.

\end{document}